\documentclass[final,12pt,times,twocolumn]{elsarticle}
\usepackage[margin=1.8cm]{geometry}
\usepackage[table]{xcolor}
\usepackage{bbm}
\usepackage{amsmath}
\usepackage{amsfonts}
\usepackage{amssymb}
\usepackage{graphicx}
\usepackage{lmodern}
\usepackage{hyperref}
\usepackage{dcolumn}
\usepackage{bm}
\usepackage{float}
\usepackage[T1]{fontenc}
\usepackage{subfigure}
\usepackage{array}

\begin{document}

\newcolumntype{P}[1]{>{\centering\arraybackslash}p{#1}}




\author[mymainaddress]{Maibam Ricky Devi\corref{mycorrespondingauthor}}
\cortext[mycorrespondingauthor]{Corresponding author}
\ead{deviricky@gmail.com}
\title{Retrieving texture zeros in 3+1 active-sterile neutrino framework under the action of $ A_{4} $ modular-invariants}
\address[mymainaddress]{Department of Physics, Gauhati University, Guwahati - 781014, Assam, India}
\begin{abstract}
The flavour problem and the viability of texture zeros of the Majorana mass matrix in the 3+1 active-sterile neutrino mixing scenario are investigated in this work using a novel bottom-up technique where we leverage the pertinent concepts of full modular group, modular invariants, and $A_4$ flavour symmetry as theoretical tools for the explicit construction of neutrino models.  In this approach we treat each chiral field as modular forms in the 3+1 neutrino mixing which are constrained by the $ A_{4} $ modular symmetry. Using these techniques, we create straightforward predictive models that only depends on a few parameters and simultaneously explains the observed pattern of neutrino mixing without the need for fine-tuning, allowing us to perceive the feasible zero textures of the Majorana mass matrix under the 3+1 framework.We discuss the implications of the allowed 3+1 zero textures by analyzing the sterile neutrino parameters, providing insight into the flavor problem and the viability of the Majorana mass matrix.
\end{abstract}

\maketitle
\section{Introduction}
\label{sec:1}
For decades, researchers have been intensely studying neutrino physics following the groundbreaking discovery of neutrino oscillation by the Super-Kamiokande experiment, which enabled the determination of the pattern of neutrino mixing. Despite the progress made in recent years, there is still much to be learned about neutrino flavor, prompting new efforts to find more effective solutions to the flavour problem. The challenge lies in detecting neutrinos, as they have very weak interactions with ordinary matter, and accurately measuring the properties of neutrino oscillations. Through advances in particle detectors and experimental techniques, researchers are continuing to make strides in understanding the flavor problem in neutrino physics.\\ 
\\
In the field of neutrino physics, physicists are currently grappling with two important issues: the flavour problem in the Standard model and the search for a fourth flavour neutrino, known as the sterile neutrino, and how it interacts with other leptons and scalar fields. To address these issues, we have developed a novel bottom-up approach that combines modular symmetries and modular forms with the $A_{4}$ flavour symmetry \cite{Feruglio:2017spp}. By implementing this $A_{4}$ modular symmetry, we can study the different texture zeros in the 3+1 neutrino mass matrices of active-sterile neutrino mixing. Incorporating the $A_{4}$ modular symmetry into the allowed texture zeros of the 3+1 neutrino mass matrix can be an extremely powerful tool in identifying the underlying mixing pattern and intricate symmetries in neutrino. It provides a framework for building elegant and efficient neutrino models with very few parameters and rarely any need of  fine-tuning. By utilizing modular symmetry, physicists can predict the properties and behavior of neutrinos with a high degree of accuracy, and modular forms offer a systematic and flexible way to build these models. 
\\
\\
One of the main advantages of using this modular invariance approach to build neutrino models is that the modular symmetry acts as the source of flavour symmetry. Instead of using the VEV of the scalar fields -flavons, to break flavour groups in the conventional flavour space, we use the modulus $\tau$ for flavour symmetry breaking. This approach offers a new perspective on the flavour problem and opens up new avenues for research in the field of neutrino physics.
\\
\\
The use of different texture zeros in neutrino mass matrices has been extensively studied as a potential solution to the flavor problem in the Standard Model. In the 3+1 framework, the allowed texture-zero patterns of lepton mass matrices have been elegantly explained and summarized in \cite{Borah:2017azf}. To further understand the behaviour of the modular forms acting on the chiral fields, we carefully examine which texture zeros( mentioned in \cite{Borah:2017azf} and shown in \ref{appen1}) would be viable under the action of $ A_{4} $ modular invariance.  We systematically analyze these texture-zero patterns for different neutrino mass models that can fit the experimental data that is based on the assumption that the fermionic leptons and scalar fields transform as irreducible representations of $ A_{4} $ \cite{Ma:2001dn}, \cite{Babu:2002dz}, \cite{Altarelli:2005yp}, \cite{Altarelli:2005yp}, \cite{Ishimori:2010au}. By exploring the allowed texture zeros in neutrino mass matrices in active-sterile neutrino mixing, we aim to provide a deeper understanding of the flavor problem in the Standard Model. Our analysis takes into account the latest experimental data and provides valuable insights into the behavior and interactions of the modular forms for level N=3 which can be transformed to a finite flavor group $ A_{4} $.\\
\\
The concept of modular symmetry has gained significant attention in global supersymmetry theories, particularly in addressing the flavor problem. To tackle this pressing issue, researchers initially explored groups such as $\Gamma_{2}\cong S_{3}$ ~\cite{Kobayashi:2018vbk,Kobayashi:2018wkl,Kobayashi:2019rzp,Okada:2019xqk}, $\Gamma_{3}\cong A_{4}$ \cite{Feruglio:2017spp,Criado:2018thu,Kobayashi:2018vbk,Kobayashi:2018scp,deAnda:2018ecu,Okada:2018yrn,Kobayashi:2018wkl,Novichkov:2018yse,Nomura:2019jxj,Okada:2019uoy,Nomura:2019yft,Ding:2019zxk,Okada:2019mjf,Nomura:2019lnr,Kobayashi:2019xvz,Asaka:2019vev,Gui-JunDing:2019wap,Zhang:2019ngf,Nomura:2019xsb,Wang:2019xbo,Kobayashi:2019gtp,King:2020qaj,Ding:2020yen,Okada:2020rjb,Nomura:2020opk,Okada:2020brs,Yao:2020qyy,Feruglio:2021dte},,  and $\Gamma_{4}\cong S_{4}$~\cite{Penedo:2018nmg,Novichkov:2018ovf,deMedeirosVarzielas:2019cyj,Kobayashi:2019mna,King:2019vhv,Criado:2019tzk,Wang:2019ovr,Gui-JunDing:2019wap,Wang:2020dbp}. As further research was conducted, other groups such as $\Gamma_{5}\cong A_{5}$~\cite{Novichkov:2018nkm,Ding:2019xna,Criado:2019tzk}, $\Gamma_{7}\cong PSL(2, Z_{7})$~\cite{Ding:2020msi}, $\Gamma^{\prime}_{3}\cong T^{\prime}$~\cite{Liu:2019khw,Lu:2019vgm}, $\Gamma^{\prime}_{4}\cong S^{\prime}_{4}$~\cite{Novichkov:2020eep,Liu:2020akv} and $ \Gamma^{\prime}_{5} \cong A^{\prime}_{5} $ ~\cite{Wang:2020lxk,Yao:2020zml,  Qu:2021jdy} were also utilized to construct models that study quark and lepton flavors and their unification. Yet, because to its infrequent fine tuning and highly predictable findings on neutrino observables, this formalism has recently attracted increased interest in the construction of neutrino models.\\
\\
Several recent studies on modular forms include \cite{Abe:2023ylh}, \cite{Ding:2023ynd}, \cite{Kim:2023jto}, \cite{Kikuchi:2023jap}, \cite{Downing:2023uuc}, \cite{Mishra:2023cjc}. Theoretically, different phenomenological ansatz, such as texture zeroes, hybrid textures, scaling matrices, and magic symmetry, to name a few, have been proposed to increase the predictability and reduce the arbitrariness of free parameters in the neutrino mass matrix. The realization of texture zeroes, hybrid textures, broken scaling, broken $\mu-\tau$, and magic symmetric neutrino models based on group symmetries has been explored to understand the possible origin of such proposed ansatz and predictions on neutrino masses and mixing \cite{Kashav:2022kpk}, \cite{Kashav:2021zir}, \cite{Kashav:2022efq}, \cite{Singh:2022tvz}, \cite{Arora:2022uof}, \cite{Arora:2022hza},  \cite{Singh:2022ijf}, \cite{Ankush:2021opd}, \cite{Verma:2021dmm}, \cite{Verma:2020gpl}, \cite{Verma:2019uiu} and \cite{Verma:2018lro}. Several recent models of neutrinos based on $A_4$ and $A_{5^{\prime}}$ modular symmetry groups  are discussed in the literature \cite{Behera:2020sfe}, \cite{Behera:2020lpd}, \cite{Behera:2021eut}, \cite{Behera:2022wco} and \cite{Mishra:2022egy}. These models explore various topics such as neutrino phenomenology, dark matter, leptogenesis, and lepton flavor violations in different seesaw mechanisms. One can refer to the references \cite{Ding:2022nzn}, \cite{Liu:2021gwa}, \cite{Li:2021buv}, \cite{Ding:2021iqp}, \cite{Ding:2020zxw}, \cite{Liu:2020msy},\cite{Nomura:2023dgk}, \cite{Benes:2022bbg}, \cite{Okada:2022kee}, \cite{Nomura:2022boj}, \cite{Ishiguro:2022pde}, \cite{Nomura:2022hxs}, \cite{Nomura:2021pld}, \cite{Dasgupta:2021ggp}, \cite{Okada:2021aoi} to gain insight on some highly sophisticated neutrino models based on modular symmetry. 
\section{$ A_{4} $ modular group of integral weight and level 3}
\label{sec:2}

We define the  full modular group $ SL_{2}(\mathbb{Z}) $ (sometimes  simply called as the modular group) as a group of $ 2\times 2 $ matrices with integer entries and unit determinant. Mathematically it can be written in the set builder form as (\textit{See ref ~ Modular Forms: A Classical and Computational Introduction by Lloyd J. P. Kilford})
\begin{equation}
\begin{aligned}
 SL_{2}(\mathbb{Z}) : = \Bigl\{
 \left(\begin{matrix}
a & b \\ c & d
\end{matrix}\right) \Big| \quad a,b,c,d \in \mathbb{Z}, && \\ ad - bc = 1 \Bigr\} ,
\end{aligned}
\label{SL2Z}
\end{equation}
The Special Linear group, $ SL_{2}(\mathbb{Z}) $ is usually denoted as $ \Gamma $ , which is a homogeneous modular group. If N is a positive integer, then the series of infinite normal subgroups of $ \Gamma $ can  be defined as principal congruence subgroups of level  N $ (N \geq 2)$. Some congruence subgroups, $ \Gamma_{0}(N)$, $ \Gamma_{1}(N)$, $ \Gamma (N)$ of  $ SL_{2}(\mathbb{Z}) $ can be written as 
\begin{equation}
\begin{aligned}
\Gamma_{0}(N) := \Bigl\{
 \left(\begin{matrix}
a & b \\ c & d
\end{matrix}\right)\in SL_{2}(\mathbb{Z}) && : && c \equiv 0 &&  mod\;  N \Bigr\} ,
\end{aligned}
\label{Gam:0}
\end{equation}
\begin{equation}
\begin{aligned}
\Gamma_{1}(N) := \Bigl\{
 \left(\begin{matrix}
a & b \\ c & d
\end{matrix}\right)\in SL_{2}(\mathbb{Z}) : && a\equiv d \equiv 1 &&  mod\; N, \\   c \equiv 0 &&  mod \; N \Bigr\} ,
\end{aligned}
\label{Gam:1}
\end{equation}
\begin{equation}
\begin{aligned}
\Gamma(N) := \Bigl\{
 \left(\begin{matrix}
a & b \\ c & d
\end{matrix}\right)\in SL_{2}(\mathbb{Z}) : && a\equiv d \equiv 1 &&  mod \; N, \\   b\equiv c \equiv 0 &&  mod \; N \Bigr\} ,
\end{aligned}
\label{Gam:N}
\end{equation}
If $ \mathbb{H} $ represents the upper half of the complex plane such that $ \mathbb{H}=Im(\mathbb{\tau})>0  $ with $ \tau $ as the complex modulus, then $\Gamma (N)$ acts on $ \mathbb{H} $ using the the linear fractional transformation given by 
\begin{eqnarray}
\gamma =  \left(\begin{matrix}
a & b \\ c & d
\end{matrix}\right) \in \Gamma : ~ \tau \rightarrow \gamma\tau = \frac{a\tau + b}{c\tau + d} \;,
\label{MG:tran}
\end{eqnarray}
The compactified space of the quotient group $ \mathbb{H}/\Gamma (N) $ is obtained by adjoing all the cusp forms at $ i\infty $ or  rational  real numbers. Apart  from  $ \Gamma $ we have the 
 infinite modular group $\overline{\Gamma}$ which have the form :

\begin{equation}
\begin{aligned}
\overline{\Gamma} = \Bigl\{
 \left(\begin{matrix}
a & b \\ c & d
\end{matrix}\right)/(\pm 1) \quad a,b,c,d \in \mathbb{Z}, && \\ ad - bc = 1 \Bigr\} ,
\end{aligned}
\label{MG:def}
\end{equation}
where $\mathbb{Z}$ refers to the set of integers. Thus from \ref{SL2Z} we can imply that $\overline{\Gamma}\equiv \Gamma/{\pm 1} $ which makes it an inhomogeneous modular group. Two generators, designated as $S$ and $T$, meet the equation$S^2 = (ST)^3 = \mathbbm{1}$ for the modular group.
The matrices for the generators $S$ and $T$ are 
\begin{eqnarray}
S = \left(\begin{matrix} 0 & 1 \\ -1 & 0 \end{matrix}\right) \;,\quad T = \left(\begin{matrix} 1 & 1 \\ 0 & 1 \end{matrix}\right)
\label{MG:rep}
\end{eqnarray}
Under $S$ and $T$, the complex modulus $\tau$ transforms as
\begin{eqnarray}
S:~ \tau \rightarrow - \frac{1}{\tau} \;,\quad T:~ \tau \rightarrow \tau + 1 \;,
\label{MG:gtran}
\end{eqnarray}
\begin{figure}[h]  
     \centering  
   \includegraphics[width=0.45\textwidth]{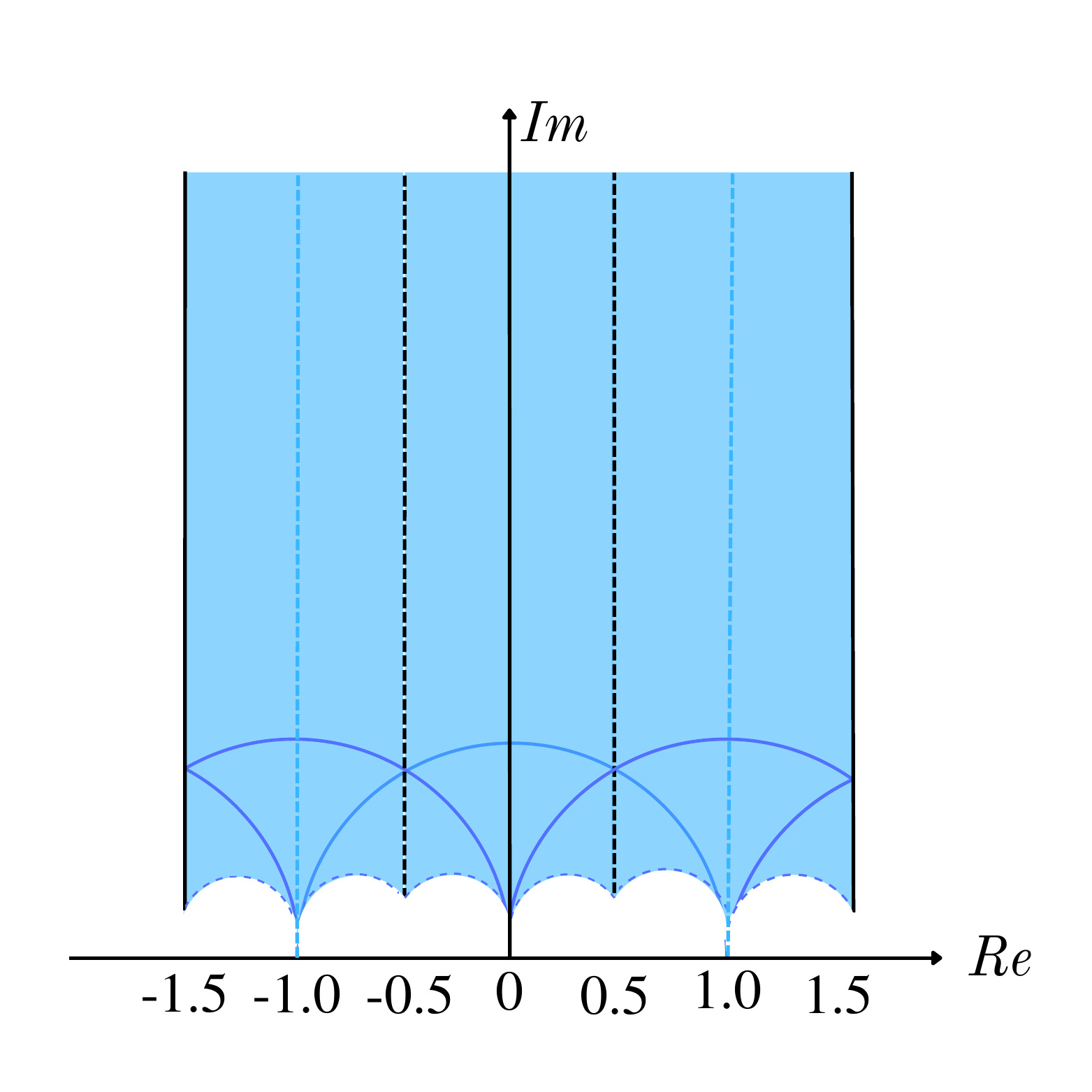} 
    \caption{}
         \label{Fig:1}
     \end{figure} 
    respectively.
    \\
    \\
 The finite modular group, denoted as $\Gamma_N$, is the quotient group of the modular group $\overline{\Gamma}$ by $\overline{\Gamma}(N)$. This group is characterized by the condition that $T^{N} = \mathbbm{1}$ holds, except for the relation $S^2 = (ST)^3 = \mathbbm{1}$. For specific values of $N$, namely $N=2,3,4,5$, the finite modular group is isomorphic to certain well-known groups. Specifically, we have $\Gamma_2 \simeq S_3$, $\Gamma_3 \simeq A_4$, $\Gamma_4 \simeq S_4$, and $\Gamma_5 \simeq A_5$. 
 \\
 \\
In  this work we primarily use modular forms of level 3 that satisfies  
\begin{equation}
\begin{aligned}
f(\gamma \tau)= (c\tau+d)^{2k}f(\tau) && \gamma =\left(\begin{matrix}
a & b \\ c & d
\end{matrix}\right)\in \Gamma(3)
\end{aligned}
\label{Gam:2}
\end{equation}
\begin{equation}
\begin{aligned}
\Gamma(3) := \Bigl\{
 \left(\begin{matrix}
a & b \\ c & d
\end{matrix}\right)\in SL_{2}(\mathbb{Z}), \;\quad\; \\ \left(\begin{matrix}
a & b \\ c & d
\end{matrix}\right) \equiv \left(\begin{matrix}
1 & 0 \\ 0 & 1
\end{matrix}\right) (mod \; 3) \Bigr\} ,
\end{aligned}
\label{Gam:3}
\end{equation}
The fundamental domain of $ \overline{\Gamma}(3) $, $ \mathbb{D} $ is given by the quotient space $ \mathbb{H}/\Gamma(3) $ which is illustrated in Fig \ref{Fig:1}.

The fundamental domain of a congruence subgroup $ \overline{\Gamma} (N) $ is defined as $ \mathbb{H} /\overline{\Gamma} (N) $, which is a quotient space. It's important to note that the fundamental domain $  \mathcal{F}$ is congruent to $ \mathbb{H} /\Gamma  $. This means that we can express $ \mathbb{H} /\overline{\Gamma} (N) $ as $ \mathbb{H} /\overline{\Gamma} (N)\equiv  \mathbb{H} /\Gamma \times \Gamma/\overline{\Gamma} (N) \equiv \mathcal{F}\times \Gamma_N$ , although this expression is less rigorous. In simpler terms, we can obtain the fundamental domain of $ \overline{\Gamma} (N)$ by applying all elements of a finite modular group $\Gamma_N$ to $  \mathcal{F}$. We transform the modular forms of level 3 undertaken in our theory based on the irreducible representations of $ A_{4} $ group. 
\\
\\
In reference \cite{Feruglio:2017spp}, the explicit construction of the basis of weight-2 modular forms of $\Gamma(3)  $ has been achieved. For those unfamiliar with the Kronecker and Tensor Product of  $A_{4}$, please refer to \ref{appen2}. These modular forms provide a triplet representation of  $A_{4}$ and can be expressed using the Dedekind eta function $ \eta (\tau) $ and its derivative. \\
\\
The Dedekind's eta-function is a fundamental example of infinite product representation which is also a holomorphic function defined in the complex upper half-plane, denoted as $ \eta(\tau) $. Thus it is clear from this definition that $ \eta(\tau) $ is non-vanishing in the complex upper half-plane $ \mathbb{H} $.
\begin{eqnarray}
\eta(\tau) = q^{1/24} \prod^{\infty}_{n=1} (1- q^n)
\label{MF:eta}
\end{eqnarray}
with $q = e^{i2\pi\tau}$. The Dedekind   eta-function and its derivative are used to find the three linearly independent modular forms of weight 2 and level 3 as ~\cite{Feruglio:2017spp}

\begin{equation}
\begin{aligned}
Y^{(2)}_{1}(\tau) = \frac{i}{2\pi} \Bigl[ \frac{\eta^{\prime}(\tau/3)}{\eta(\tau/3)}  + \frac{\eta^{\prime}((\tau+1)/3)}{\eta((\tau+1)/3)} \\   +\frac{\eta^{\prime}((\tau+2)/3)}{\eta((\tau+2)/3)}  -\frac{27\eta^{\prime}(3\tau)}{\eta(3\tau)} \Bigr] \;,
\\
Y^{(2)}_{2}(\tau) = -\frac{i}{\pi} \Bigl[ \frac{\eta^{\prime}(\tau/3)}{\eta(\tau/3)} + \omega^2 \frac{\eta^{\prime}((\tau+1)/3)}{\eta((\tau+1)/3)} \\   + \omega \frac{\eta^{\prime}((\tau+2)/3)}{\eta((\tau+2)/3)} \Bigr] \;,
\\
Y^{(2)}_{3}(\tau) = -\frac{i}{\pi} \Bigl[ \frac{\eta^{\prime}(\tau/3)}{\eta(\tau/3)} + \omega \frac{\eta^{\prime}((\tau+1)/3)}{\eta((\tau+1)/3)} \\   + \omega^2 \frac{\eta^{\prime}((\tau+2)/3)}{\eta((\tau+2)/3)} \Bigr] \;,
\label{MF:form}
\end{aligned}
\end{equation}
where $\omega=e^{i2\pi/3}$. Eqns.(\ref{MF:form}) later satisfy the constraint: ~\cite{Feruglio:2017spp}
\begin{eqnarray}
Y^{(2)2}_{2} + 2 Y^{(2)}_{1} Y^{(2)}_{3} = 0 \;.
\label{MF:cons}
\end{eqnarray}
These three modular forms can be organized into a triplet of $A_4$ transforming in the ${\bf 3 }$ irreducible representation of $A_4$, denoted as
$ Y^{(2)}_{i}(\tau)_{i=1,2,3} $ can be grouped together to transform as the triplet representation of $A_4$ as

\begin{eqnarray}
Y^{(2)}_{\bf 3} (\tau) = \left(\begin{matrix} Y^{(2)}_1(\tau) \\ Y^{(2)}_2 (\tau)\\ Y^{(2)}_3(\tau) \end{matrix}\right)\;,
\label{MF:2}
\end{eqnarray}
Following the q-expansion of $ Y_{i}(\tau) $ by considering $ q_{3} \equiv e^{\dfrac{i2\pi \tau}{3}} $, Eqn. (\ref{MF:form}) takes the form:
\begin{equation}
\begin{aligned}
Y^{(2)}_{\textbf{3}}(\tau) = \left( \begin{matrix}
Y^{(2)}_{1}(\tau)\\
Y^{(2)}_{2}(\tau)\\
Y^{(2)}_{3}(\tau)
\end{matrix} \right)  \\ = \left( \begin{matrix}
1+12q^{3}_{3} +36q^{6}_{3} +12+36q^{9}_{3}+...\\
-6(q^{}_{3} + 7q^{4}_{3}+8q^{7}_{3}+...)\\
-18(q^{2}_{3}+ 2q^{5}_{3}+5q^{8}_{3}+...)
\end{matrix} \right)  \;.
\label{MF:form2}
\end{aligned}
\end{equation}
It is possible to select a basis in the linear space of $ M_{2k} (\Gamma(N))$  with weight 2k such that the transformation of the modular forms $Y(\tau)$ is defined by the unitary representation  $ \rho $ of $ \Gamma_N $. Modular multiplets $Y(\tau)$ with a weight of 2k and a level of $ N\geq 2$ yield non-vanishing modular forms that are consistently of even weight~\cite{Feruglio:2017spp}. This unique property is particularly noteworthy as it allows for greater flexibility in the analysis of modular forms. Furthermore, the function does not vanish for odd weight modular forms, further enhancing the versatility of this approach.

\begin{eqnarray}
Y_i(\gamma \tau) = (c\tau + d)^k \rho(\gamma)_{ij} Y_j(\tau) \;,
\label{MF:tran}
\end{eqnarray}
The earliest discussion of modular forms of general positive integer weights (including positive even weights and odd weights) and finite modular groups can be seen in the literature \cite{Liu:2019khw}, which also contains the proof of Eqn. (\ref{MF:tran}), which is very helpful for understanding modular form multiplets.
Modular forms are crucial in the construction of models to explain fermion masses and flavour mixing ~\cite{Feruglio:2017spp}. The linear space that is covered by modular forms of weight $k$ has a dimension of $k+1$ for the case of level 3 that corresponds to the group  $\Gamma_3 \simeq A_4$. 
Thus we can obtain only one trivial modular form, which is constant and independent of the modulus $\tau$, for $k=0$. A triplet of $A_4$ is formed for $k=2$ by three linearly independent modular forms.
\\
\\
 Using the tensor product rules of $A_4$  as shown in \ref{appen2}, we can construct the higher weight modular forms ( $ k\geq 4 $ here) with the help of Eqs.~(\ref{MF:cons}) and (\ref{MF:2}. In our models we use $ k_{Y_{max}} =4,6,8,10 $. For weight $k=4$, we have  two singlets ${\bf 1}$, ${\bf 1^\prime}$ and one triplet ${\bf 3}$ of $A_4$ with a total of 5 linearly independent modular forms:
\begin{equation}
\begin{aligned}
Y^{(4)}_{\bf 1} = Y^2_1 + 2Y_2Y_3  \;, \quad  Y^{(4)}_{\bf 1^\prime} = Y^2_3 + 2Y_1Y_2 \;, \quad  \\  
 Y^{(4)}_{\bf 3} = \left(\begin{matrix} Y^2_1 - Y_2Y_3 \\ Y^2_3 - Y_1Y_2 \\ Y^2_2 - Y_1Y_3  \end{matrix} \right) \;
\end{aligned}
\label{MF:4}
\end{equation}
Similarly when $k=6$, one has 
\begin{equation}
\begin{aligned}
Y^{(6)}_{\bf 1} = Y^3_1 + Y^3_2 + Y^3_3 -3Y_1Y_2Y_3 \;,  \\ 
Y^{(6)}_{{\bf 3},1} = (Y^2_1 + 2Y_2Y_3) \left(\begin{matrix} Y_1 \\ Y_2 \\ Y_3 \end{matrix}\right) \;,  \quad  \\   Y^{(6)}_{{\bf 3},2} = (Y^2_3 + 2Y_1Y_2) \left(\begin{matrix} Y_3 \\ Y_1 \\ Y_2 \end{matrix}\right)
\end{aligned}
\label{MF:6}
\end{equation}
the dimension is 7 where $ Y^{(3)}_{3,3}=0 $. Following the same procedure, for $k=8$ , we have
\begin{equation}
\begin{aligned}
Y^{(8)}_{\bf 1} = (Y^2_1 + 2Y_2Y_3)^2 \;,\quad  \\  Y^{(8)}_{\bf 1^\prime} = (Y^2_1 + 2Y_2Y_3) (Y^2_3 + 2Y_1Y_2) \;,\quad \\  Y^{(8)}_{\bf 1^{\prime\prime}} = (Y^2_3 + 2Y_1Y_2)^2 \;,
 \\ 
Y^{(8)}_{{\bf 3},1} = (Y^2_1 + 2Y_2Y_3) \left(\begin{matrix} Y^2_1 - Y_2Y_3 \\ Y^2_3 - Y_1Y_2 \\ Y^2_2 - Y_1Y_3 \end{matrix}\right) \;,\quad  \\ Y^{(8)}_{{\bf 3},2} = (Y^2_3 + 2Y_1Y_2) \left(\begin{matrix} Y^2_2 - Y_1Y_3 \\ Y^2_1 -Y_2Y_3 \\ Y^2_3 - Y_1Y_2 \end{matrix}\right) \;, 
\end{aligned}
\label{MF:8}
\end{equation}
a collective dimension of 9. For $k=10$, we get  11 linearly independent modular forms which  are assigned as  $A_4$ multiplets as  shown below
\begin{equation}
\begin{aligned}
Y^{(10)}_{\bf 1} = (Y^2_1 + 2Y_2Y_3) (Y^3_1 + Y^3_2 + Y^3_3 - 3Y_1Y_2Y_3) \;,
\\
Y^{(10)}_{\bf 1^\prime} = (Y^2_3 + 2Y_1Y_2) (Y^3_1 + Y^3_2 + Y^3_3 - 3Y_1Y_2Y_3) \;,
\\
Y^{(10)}_{{\bf 3},1} = (Y^2_1 + 2Y_2Y_3)^2 \left(\begin{matrix} Y_1 \\ Y_2 \\ Y_3 \end{matrix}\right) , \\  Y^{(10)}_{{\bf 3},2} =  (Y^2_3 + 2Y_1Y_2)^2 \left(\begin{matrix} Y_2 \\ Y_3 \\ Y_1 \end{matrix}\right) \;,
\\
Y^{(10)}_{{\bf 3},3} = (Y^2_1 + 2Y_2Y_3) (Y^2_3 + 2Y_1Y_2) \left(\begin{matrix} Y_3 \\ Y_1 \\ Y_2 \end{matrix}\right) \;.
\end{aligned} 
\label{MF:10}
\end{equation}
These modular forms of higher weights are shown explicitly in ~\cite{Feruglio:2017spp} and \cite{Zhang:2019ngf}.
The generic form of matter action in the Lagrangian $\mathcal N = 1$ global supersymmetry theory  is ~\cite{Ferrara:1989bc,Ferrara:1989qb}
\begin{eqnarray}
{\mathcal S} = \int {\rm d}^4x {\rm d}^2\theta {\rm d}^2\overline{\theta} K( \phi_i, \overline{\phi}_i; \tau, \overline{\tau}) \\\nonumber + \int {\rm d}^4x {\rm d}^2\theta W(\phi_i; \tau) + {\rm h.c.} \;,
\label{Action}
\end{eqnarray}
Here $\theta  $ and $ \overline{\theta} $ represents the Grassmann variables, $\phi_i$ is the collective matter chiral superfields involved in the theory and K and W stands for the K$\ddot{\rm a}$hler potential and superpotential respectively.  the modulus  $\tau$ as well as $\phi_i$ undergo transformation as shown in Eq.~(\ref{MG:tran}) and Eq. (\ref{MG:csft}) ~\cite{Ferrara:1989bc} for a finite modular group $\Gamma_N$ (to the limit of a field - independent phase $ \alpha (\gamma) $ known as the multiplier system) \cite{Ferrara:1989qb}, \cite{Ferrara:1989bc}

\begin{eqnarray}
\gamma:~ \phi^{}_i \rightarrow (c\tau + d)^{-k_i}_{} \rho^{}_{I_i}(\gamma) \phi^{}_i \;,
\label{MG:csft}
\end{eqnarray} 
From this, the superpotential $W(\phi_i;\tau)$ can be expanded in powers of the supermultiplets $\phi_i$ as
\begin{eqnarray}
W(\phi_i,\tau) = \sum (Y_{i_1,\cdots,i_n}(\tau) \phi_{i_1} \cdots \phi_{i_n})^{}_{\bf 1} \;,
\label{Sp}
\end{eqnarray}
where ${\bf 1}$ denotes the sum of all possible field combinations and singlets of the modular group $\Gamma_N$. Inorder for the superpotential to under field transformation as the rule given in 
 Eq.~(\ref{MG:csft}), $Y^{(k_Y)}_{I_Y}$ being the field couplings need to transform as modular forms of  weight $k_Y$ and level N such that $k_Y = k_{i_1} + \cdots + k_{i_n}$ holds under $ \Gamma_{N}$, i.e.,
\begin{eqnarray}
\gamma: ~ Y^{(k_Y)}_{I_Y} (\tau) \rightarrow Y^{(k_Y)}_{I_Y} (\gamma\tau) = (c\tau + d)^{k_Y} \\ \nonumber \times \rho^{}_{I_Y}(\gamma)  Y^{(k_Y)}_{I_Y} (\tau) \;.
\label{MG:mf}
\end{eqnarray}
The unitary representation of $ \Gamma_{N}$, $ \rho $ must satisfy    $\rho_{I_Y} \otimes \rho_{I_1} \otimes \cdots \otimes \rho_{I_n}$ such that it involved atleast one singlet invariant under the  finite modular group $ \Gamma_{N}$.

\section{$ A_{4} $ modular invariant zero-textures in 3+1 framework using minimal seesaw mechanism}
\label{sec:3}
In the context of a diagonal charged lepton mass matrix $ M_{l} $, there exist fifteen distinct two-zero textures, with twenty out of 120 allowed three-zero textures, fifteen out of 210 allowed four-zero textures, and six out of 246 allowed five-zero textures in the Majorana neutrino mass matrix $ m_{\nu} $. These findings are detailed in the reference. Additionally, \ref{appen1} briefly outlines the various zero textures that are not ruled out, each of which is labeled with a specific alphabet and numeric designation where in the matrices of the possible textures "$ \times $" represents a non-zero entry.\\
\\
In our work we have realized these possible textures using the non-Abelian $ A_{4} $ flavour symmetry. We have incorporated modular symmetry along with the $ A_{4} $ flavour symmetry to reduce the number of free parameters and to create a more predictable models without fine tuning. The reason we chose $ A_{4} $ flavour symmetry is because it has three irreducible singlet representations and one triplet representation that can naturally accomodate its triplet family of neutrinos and its three right-handed singlets. Here we generate the neutrino masses using minimal extended seesaw mechanism. The neutral fermionic mass matrix can be written in the basis $ (L, N, S)$ as 
\begin{equation}
\begin{aligned}
\mathbb{M}=\left( \begin{matrix}
0 & M_{D} & 0\\
M^{T}_{D} & M_{R} & M^{T}_{S}\\
0 & M_{S} & 0
\end{matrix} \right)
\end{aligned}
\end{equation}
\begin{table}[h]
	\centering
	\vspace{-0.2cm}
	\caption{A typical assignment of representations and weights for the chiral supermultiplets in Cases (a), (b), (c) and (d).}
	\setlength{\tabcolsep}{0.1cm}
	\vspace{0.3cm}
	\begin{tabular}{|c|c|c|c|c|c|c|c|c|c|c|c|c|}
		\hline
		& $L_1$ & $L_2$ & $L_3$ & $E^c_1$ & $E^c_2$ & $E^c_3$ & $H_u$ & $H_d$ & $N^{c}_{1}$ & $N^{c}_{2}$ & $S$ & $ \phi $\\
		\hline
		$A_4$ & ${\bf 1}$ &  ${\bf 1^{\prime \prime}}$ &  ${\bf 1^{\prime}}$ &  ${\bf 1}$ &  ${\bf 1^{\prime}}$ &  ${\bf 1^{\prime\prime}}$ &  ${\bf 1}$ &  ${\bf 1}$ & ${\bf 1^\prime}$ & ${\bf 1^{\prime\prime}}$ & ${\bf 1^{\prime}}$ & 1\\
		\hline
		$-k$ & $k_1$ & $k_2$ & $k_3$ &  $r_1$ & $r_2$ & $r_3$ & 0 & 0 & $l_1$ & $l_2$ & $s$ & $ k_{\phi} $ \\
		\hline
	\end{tabular}
	\label{cont1}
\end{table}
\begin{table}[h]
	\centering
	\vspace{-0.2cm}
	\caption{A typical assignment of representations and weights for the chiral supermultiplets in Cases (e) and (f).}
	\setlength{\tabcolsep}{0.1cm}
	\vspace{0.3cm}
	\begin{tabular}{|c|c|c|c|c|c|c|c|c|c|c|c|c|}
		\hline
		& $L_1$ & $L_2$ & $L_3$ & $E^c_1$ & $E^c_2$ & $E^c_3$ & $H_u$ & $H_d$ & $N^{c}_{1}$ & $N^{c}_{2}$ & $S$ & $ \phi $\\
		\hline
		$A_4$ & ${\bf 1}$ &  ${\bf 1^\prime}$ &  ${\bf 1^{\prime\prime}}$ &  ${\bf 1}$ &  ${\bf 1^{\prime\prime}}$ &  ${\bf 1^\prime}$ &  ${\bf 1}$ &  ${\bf 1}$ & ${\bf 1}$ & ${\bf 1^\prime}$ & ${\bf 1^{\prime\prime}}$ & $ 1^{\prime \prime} $\\
		\hline
		$-k$ & $k_1$ & $k_2$ & $k_3$ &  $r_1$ & $r_2$ & $r_3$ & 0 & 0 & $l_1$ & $l_2$ & $s$ & $ k_{\phi}  $  \\
		\hline
	\end{tabular}
	\label{cont2}
\end{table}
The Dirac mass matrix  with basis $ (L, N) $ is given as
\begin{equation}
 M_{D}=\left(\begin{matrix}
a_{11} & a_{12} \\
a_{21} & a_{22} \\
a_{31} & a_{32} 
\end{matrix}\right) 
\label{ISS: MD}
\end{equation}
where $ L $ is the  left-handed lepton doublet with three generations, $ E^{c} $ is the right-handed charged leptons of three families, $ N^{c} $ is the two right-handed neutrinos, and S is the sterile neutrino. $ H_{u,d} $ defines the Higgs doublets with zero weight.
The rest of the two matrices can be written as 
\begin{equation}
  M_{R}=\left(\begin{matrix}
0 & M_{N} \\
M_{N} & 0 
\end{matrix}\right)
\label{ISS:M}
\end{equation}
\begin{equation}
 M_{S}=\left(\begin{matrix}
\rho_{1} & \rho_{2} 
\end{matrix}\right) 
\label{ISS:M_s}
\end{equation}
If the condition $ M_{R} \gg M_{S}> M_{D} $ then for minimal seesaw mechanism, the effective  $  4 \times 4$ active-sterile neutrino mass matrix is  given by
\begin{equation}
  m_{\nu}=-\left(\begin{matrix}
M_{D}M^{-1}_{R}M^{T}_{D} & M_{D}M^{-1}_{R}M^{T}_{S}\\
M_{S}(M^{-1}_{R})^{T}M^{T}_{D} & M_{S}M^{-1}_{R}M^{T}_{S} 
\end{matrix}\right) 
\label{ISS:M_nu}
\end{equation}
We have to take care of the fact that atleast of the neutrino mass eigenvalue is zero. Eqn. (\ref{ISS:M_nu}) thus takes the form:
\begin{equation}
m_{\nu}=
\left(
\begin{matrix}
p_{1} & q & r & s\\
q & p_{2} & m & t\\
r & m & p_{3} & w\\
s & t & w & p_{4} 
\end{matrix}\right)
\label{eqn:nu}
\end{equation}
where $ p_{1} = \frac{2 a_{11} a_{12}}{M_N} $, $ p_{2} = \frac{2 a_{12} a_{22}}{M_N} $, $ p_{3} = \frac{2 a_{31} a_{32}}{M_N} $, $ p_{4} = \frac{2 \rho_1 \rho_2}{M_N}$, $ q=\frac{a_{12}^2}{M_N}+\frac{a_{11} a_{22}}{M_N} $, $ r= \frac{a_{12} a_{31}}{M_N}+\frac{a_{11} a_{32}}{M_N} $, $ s= \frac{a_{12} \rho_1}{M_N}+\frac{a_{11} \rho_2}{M_N} $, $m= \frac{a_{22} a_{31}}{M_N}+\frac{a_{12} a_{32}}{M_N}$, $ t=\frac{a_{22} \rho_1}{M_N}+\frac{a_{12} \rho_2}{M_N}  $, $ w=\frac{a_{32} \rho_1}{M_N}+\frac{a_{31} \rho_2}{M_N} $.

To simplify our analysis, we have included a table (\ref{cont}) that displays the $A_{4}$ representations and modular weights assigned to the chiral supermultiplets. In our models, we have established fixed textures for $M_{R}$ and $M_{S}$ that are invariant under the $A_{4}$ modular symmetry.  However, the textures of $M_{D}$ do transform along with the assigned weights and $A_{4}$ representations. We then apply minimal seesaw mechanism to find only the allowed textures out of the possible textures as shown in \ref{appen1} under the $ A_{4} $ modular symmetry. It is interesting to find that out of all the possible cases only 6 textures (3 two-zeros+ 1 three-zeros + 2 four-zeros) are allowed and the rest are ruled out which is summarized in Table (\ref{tab:1}). These models are shown below where $ \alpha $, $ \beta $, $ \gamma $, $ \lambda_{ij} $ (i= 1,2,3; j=1,2), $\Omega$, $ \rho $ are arbritary coefficients with $ \langle H_{u} \rangle = \upsilon_{u}$ and there is absence of non-trivial modular forms in the charged lepton sector. For cases (a), (b), (c),  and (d), we are following Table (\ref{cont1}) and for cases (e) and (f) we are following Table (\ref{cont2}) where the particle content of the various fields under $A_{4}$ modular transformation is given.
\begin{table}[h]
        \centering
         \scalebox{0.573}{
        \begin{tabular}{| c | c | c | c |c|}
        \hline 
\cline{1-4}
\cellcolor{gray!20} Case No. & \cellcolor{gray!20} Constraints & \cellcolor{gray!20} $ M_{D} $  & \cellcolor{gray!20} $ m_{\nu}  $ & \cellcolor{gray!20} Allowed textures\\ 
\hline
 (a) & $ a_{11}=a_{22}=0 $ &  $ \left(\begin{matrix}
 0 & \times \\
\times & 0 \\
\times & \times 
\end{matrix}\right) $& $ \left(\begin{matrix}
0 & \times & \times & \times\\
\times & 0 & \times & \times\\
\times & \times & \times &\times\\
\times & \times & \times &\times\\
\end{matrix}\right) $   & $  E_{1}$ of two-zeros \\
\hline 
(b) &$ a_{11}=a_{32}=0 $ & $ \left(\begin{matrix}
 0 & \times \\
\times & \times \\
\times & 0 
\end{matrix}\right) $ & $ \left(\begin{matrix}
0 & \times & \times & \times\\
\times & \times & \times & \times\\
\times & \times & 0 &\times\\
\times & \times & \times &\times\\
\end{matrix}\right) $ & $  E_{2}$ of two-zeros \\
 \hline
 (c) & $ a_{22}=a_{31}=0 $ & $ \left(\begin{matrix}
 \times & \times \\
\times & 0 \\
0 & \times 
\end{matrix}\right) $ & $ \left(\begin{matrix}
\times & \times & \times & \times\\
\times & 0 & \times & \times\\
\times & \times & 0 &\times\\
\times & \times & \times &\times\\
\end{matrix}\right) $ & $ C$ of two-zeros \\
 \hline
 (d) & $ a_{22}=a_{32}=0 $ & $ \left(\begin{matrix}
 \times & \times \\
\times & 0 \\
\times & 0
\end{matrix}\right) $ & $ \left(\begin{matrix}
\times & \times & \times & \times\\
\times & 0 & 0 & \times\\
\times & 0 & 0 &\times\\
\times & \times & \times &\times\\
\end{matrix}\right) $  & $ C_{3}$ of three-zeros \\
 \hline
 (e) &$ a_{12}= a_{21}=a_{31}=0 $ & $ \left(\begin{matrix}
 \times & 0 \\
0 & \times \\
0 & \times 
\end{matrix}\right) $ & $ \left(\begin{matrix}
 0 & \times & \times & \times\\
\times & 0 & 0 & \times\\
\times & 0 & 0 &\times\\
\times & \times & \times &\times\\
\end{matrix}\right) $ & $  A_{3}$ of four-zeros \\
 \hline
(f) & $ a_{12}=a_{21}=a_{32}=0 $ & $ \left(\begin{matrix}
 \times & 0 \\
0 & \times \\
\times & 0 
\end{matrix}\right) $ & $ \left(\begin{matrix}
 0 & \times & 0 & \times\\
\times & 0 & \times & \times\\
0 & \times & 0 &\times\\
\times & \times & \times &\times\\
\end{matrix}\right) $ & $  A_{2}$ of four-zeros \\
 \hline
\end{tabular}}
      \caption{Possible textures in $ m_{\nu} $ in the 3+1 scenario for different cases of $ M_{D} $ and $ m_\nu $}
        \label{tab:1}
             \end{table}

\begin{center}
(a) ${\bf E_{1}}$ of Two-zero textures
\end{center}
\begin{eqnarray}
k_1=-2\;,\; k_2=-4\;,\;k_3=-2\;, \quad \\ \nonumber r_1=2 \;,\; r_2=4\;,\;r_3=2 \;, \quad \\ \nonumber l_1 = -2\;,\; l_2 = -6 \;,\; s = -2\;, k_{\phi} = 2.
\label{case a}
\end{eqnarray}
In this model of texture $ E_{1} $ of two-zeros, we utilize the multiplets of modular forms, namely $Y^{(4)}_{1^{\prime}}$,  $Y^{(6)}_{1}$,  $Y^{(8)}_{1}$, and  $Y^{(8)}_{1^{\prime}}$, which were derived earlier. This results in a maximum value of $k_{Y{\rm max}} = 8$. The superpotential can be explicitly expressed as follows:
\begin{eqnarray}
W = &&(\alpha L^{}_1 H^{}_dE^c_1 +  \beta L^{}_2 H^{}_dE^c_2   + \gamma L^{}_3 H^{}_dE^c_3 ) 
\nonumber
\\
&& + (\lambda_{12} Y^{(8)}_{\bf 1^{\prime}} L^{}_1H^{}_u N^c_{2} + \lambda_{21} Y^{(6)}_{\bf 1}  L^{}_2H^{}_u N^c_{1}  
\nonumber
\\
&& + \lambda_{31} Y^{(4)}_{\bf 1^\prime}  L^{}_3H^{}_u N^c_{1}  + \lambda_{32} Y^{(8)}_{\bf 1}  L^{}_3H^{}_u N^c_{2}  )
\nonumber
\\
&&  + \Omega Y^{(6)}_{ \bf 1} \phi (N^c_{1}N^c_{2}+ N^c_{2}N^c_{1})
\\
&& + \rho( Y^{(4)}_{\textbf{1}^{\prime}} S N^c_{1} +  Y^{(8)}_{\textbf{1}}S N^c_{2}) \;\nonumber,
\label{casea:ea}
\end{eqnarray}
Thus from Eqns. (\ref{casea:ea}) and (\ref{eqn:nu}) we get
\begin{equation}
\resizebox{.45\textwidth}{!}
{$m_{\nu}=
\left(
\begin{matrix}
0 & \frac{a_{12}^2}{M_N}+\frac{a_{11} a_{22}}{M_N} & \frac{a_{12} a_{31}}{M_N}+\frac{a_{11} a_{32}}{M_N} &\frac{a_{12} \rho_1}{M_N}+\frac{a_{11} \rho_2}{M_N}\\
 \frac{a_{12}^2}{M_N}+\frac{a_{11} a_{22}}{M_N} & 0 & \frac{a_{22} a_{31}}{M_N}+\frac{a_{12} a_{32}}{M_N}&  \frac{a_{22} \rho_1}{M_N}+\frac{a_{12} \rho_2}{M_N} \\
 \frac{a_{12} a_{31}}{M_N}+\frac{a_{11} a_{32}}{M_N} & \frac{a_{22} a_{31}}{M_N}+\frac{a_{12} a_{32}}{M_N} & \frac{2 a_{31} a_{32}}{M_N}&\frac{a_{32} \rho_1}{M_N}+\frac{a_{31} \rho_2}{M_N} \\
  \frac{a_{12} \rho_1}{M_N}+\frac{a_{11} \rho_2}{M_N} & \frac{a_{22} \rho_1}{M_N}+\frac{a_{12} \rho_2}{M_N} & \frac{a_{32} \rho_1}{M_N}+\frac{a_{31} \rho_2}{M_N} &\frac{2 \rho_1 \rho_2}{M_N} 
\end{matrix}\right)$
}
\label{eqn:nu:a}
\end{equation}
where, $ a_{12} = \lambda_{12} Y^{(8)}_{1^{\prime }} \upsilon_{u}$, $ a_{21} = \lambda_{21} Y^{(6)}_{1} \upsilon_{u}$, $ \; \quad $ $ a_{31} = \lambda_{31} Y^{(4)}_{1^{\prime}} \upsilon_{u}$, $ a_{32} = \lambda_{32} Y^{(8)}_{1} \upsilon_{u}$, $ M_{N} = \Omega Y^{(6)}_{1} \phi$ , $ \rho_{1}= \rho Y^{(4)}_{1^{\prime}} $ and $ \rho_{2} = \rho Y^{(8)}_{1}$.

\begin{center}
(b) ${\bf E_{2}}$ of Two-zero textures
\end{center}
\begin{eqnarray}
k_1=-4\;,\; k_2=-4\;, k_3=2\;, \quad \\ \nonumber r_1=4 \;, r_2=4\;, r_3=-2\;, \quad \\ \nonumber l_1 = -6\;,\; l_2 = -4 \;,\; s = -2\;, k_{\phi}  = 6.
\label{case b}
\end{eqnarray}
In this model of texture $ E_{2} $ of two-zeros, we utilize the multiplets of modular forms, namely $Y^{(4)}_{1}$,  $Y^{(4)}_{1^{\prime}}$,  $Y^{(6)}_{1}$, $Y^{(8)}_{1^{\prime}}$, $Y^{(8)}_{1^{\prime \prime}}$, and  $Y^{(10)}_{1}$,  which were derived earlier. This results in a maximum value of $k_{Y{\rm max}} = 10$. The superpotential can be explicitly expressed as follows:
\begin{eqnarray}
W = &&(\alpha L^{}_1 H^{}_dE^c_1 +  \beta L^{}_2 H^{}_dE^c_2   + \gamma L^{}_3 H^{}_dE^c_3 ) 
\nonumber
\\
&& + (\lambda_{12} Y^{(8)}_{\bf 1^{\prime }} L^{}_1 H^{}_u N^c_{2}  + \lambda_{21} Y^{(10)}_{\bf 1} L^{}_2 H^{}_u N^c_{1}  
\nonumber
\\
&& + \lambda_{22} Y^{(8)}_{\bf 1^{\prime \prime}}  L^{}_2H^{}_u N^c_{2}  + \lambda_{31} Y^{(4)}_{\bf 1^\prime}  L^{}_3H^{}_u N^c_{1})
\nonumber
\\
&&  + \Omega Y^{(4)}_{ \bf 1} \phi (N^c_{1}N^c_{2}+ N^c_{2}N^c_{1})
\\
&& + \rho( Y^{(8)}_{\textbf{1}^{\prime}} S N^c_{1} +  Y^{(6)}_{\textbf{1}}S N^c_{2}) \;\nonumber,
\label{casea:eb}
\end{eqnarray}

\begin{equation}
\resizebox{.45\textwidth}{!}
{$m_{\nu}=
\left(
\begin{matrix}
0 & \frac{a_{12}^2}{M_N}+\frac{a_{11} a_{22}}{M_N} & \frac{a_{12} a_{31}}{M_N}+\frac{a_{11} a_{32}}{M_N} &\frac{a_{12} \rho_1}{M_N}+\frac{a_{11} \rho_2}{M_N}\\
 \frac{a_{12}^2}{M_N}+\frac{a_{11} a_{22}}{M_N} & \frac{2 a_{12} a_{22}}{M_N} & \frac{a_{22} a_{31}}{M_N}+\frac{a_{12} a_{32}}{M_N}&  \frac{a_{22} \rho_1}{M_N}+\frac{a_{12} \rho_2}{M_N} \\
 \frac{a_{12} a_{31}}{M_N}+\frac{a_{11} a_{32}}{M_N} & \frac{a_{22} a_{31}}{M_N}+\frac{a_{12} a_{32}}{M_N} & 0 &\frac{a_{32} \rho_1}{M_N}+\frac{a_{31} \rho_2}{M_N} \\
  \frac{a_{12} \rho_1}{M_N}+\frac{a_{11} \rho_2}{M_N} & \frac{a_{22} \rho_1}{M_N}+\frac{a_{12} \rho_2}{M_N} & \frac{a_{32} \rho_1}{M_N}+\frac{a_{31} \rho_2}{M_N} &\frac{2 \rho_1 \rho_2}{M_N} 
\end{matrix}\right)$
}
\label{eqn:nu:b}
\end{equation}
where, $ a_{12} = \lambda_{12} Y^{(8)}_{1^{\prime }} \upsilon_{u}$, $ a_{21} = \lambda_{21} Y^{(10)}_{1} \upsilon_{u}$, $ \; \quad  \; $    $ a_{22} = \lambda_{22} Y^{(8)}_{1^{\prime \prime}} \upsilon_{u}$,  $ a_{31} = \lambda_{31} Y^{(4)}_{1^{\prime}} \upsilon_{u}$,  $ M_{N} = \Omega Y^{(4)}_{1} \phi$, $ \rho_{1}= \rho Y^{(8)}_{1^{\prime}} $ and $ \rho_{2} = \rho Y^{(6)}_{1}$.

\begin{center}
(c) ${\bf C}$ of Two-zero textures
\end{center}
\begin{eqnarray}
k_1=-4\;,\; k_2=-4\;, k_3=-2\;, \quad \\ \nonumber r_1=4 \;, r_2=4\;, r_3=2\;, \quad \\ \nonumber l_1 = -4\;,\; l_2 = -6 \;,\; s = -4\;, k_{\phi}=6.
\label{case c}
\end{eqnarray}
In this model of texture $ C $ of two-zeros, we utilize the multiplets of modular forms, namely $Y^{(4)}_{1}$, $Y^{(8)}_{1}$, $Y^{(8)}_{1^{\prime}}$, $Y^{(8)}_{1^{\prime \prime}}$, $Y^{(10)}_{1}$, and $Y^{(10)}_{1^{\prime}}$ which were derived earlier. This results in a maximum value of $k_{Y{\rm max}} = 10$. The superpotential can be explicitly expressed as follows:
\begin{eqnarray}
W = &&(\alpha L^{}_1 H^{}_dE^c_1 +  \beta L^{}_2 H^{}_dE^c_2   + \gamma L^{}_3 H^{}_dE^c_3 ) 
\nonumber
\\
&& + (\lambda_{11} Y^{(8)}_{\bf 1^{\prime \prime}} L^{}_1 H^{}_u N^c_{1}  + \lambda_{12} Y^{(10)}_{\bf 1^{\prime}} L^{}_1 H^{}_u N^c_{2}  
\nonumber
\\
&& + \lambda_{21} Y^{(8)}_{\bf 1}  L^{}_2H^{}_u N^c_{1}  + \lambda_{32} Y^{(8)}_{\bf 1}  L^{}_3H^{}_u N^c_{2})
\nonumber
\\
&&  + \Omega Y^{(4)}_{ \bf 1} \phi (N^c_{1}N^c_{2}+ N^c_{2}N^c_{1})
\\
&& + \rho( Y^{(8)}_{\textbf{1}^{\prime}} S N^c_{1} +  Y^{(10)}_{\textbf{1}}S N^c_{2}) \;\nonumber,
\label{casea:ec}
\end{eqnarray}

\begin{equation}
\resizebox{.45\textwidth}{!}
{$m_{\nu}=
\left(
\begin{matrix}
 \frac{2 a_{11} a_{12}}{M_N} & \frac{a_{12}^2}{M_N}+\frac{a_{11} a_{22}}{M_N} & \frac{a_{12} a_{31}}{M_N}+\frac{a_{11} a_{32}}{M_N} &\frac{a_{12} \rho_1}{M_N}+\frac{a_{11} \rho_2}{M_N}\\
 \frac{a_{12}^2}{M_N}+\frac{a_{11} a_{22}}{M_N} & 0 & \frac{a_{22} a_{31}}{M_N}+\frac{a_{12} a_{32}}{M_N}&  \frac{a_{22} \rho_1}{M_N}+\frac{a_{12} \rho_2}{M_N} \\
 \frac{a_{12} a_{31}}{M_N}+\frac{a_{11} a_{32}}{M_N} & \frac{a_{22} a_{31}}{M_N}+\frac{a_{12} a_{32}}{M_N} & 0 &\frac{a_{32} \rho_1}{M_N}+\frac{a_{31} \rho_2}{M_N} \\
  \frac{a_{12} \rho_1}{M_N}+\frac{a_{11} \rho_2}{M_N} & \frac{a_{22} \rho_1}{M_N}+\frac{a_{12} \rho_2}{M_N} & \frac{a_{32} \rho_1}{M_N}+\frac{a_{31} \rho_2}{M_N} &\frac{2 \rho_1 \rho_2}{M_N} 
\end{matrix}\right)$
}
\label{eqn:nu:c}
\end{equation}
where, $ a_{11} = \lambda_{11} Y^{(8)}_{1^{\prime \prime}} \upsilon_{u}$, $ a_{12} = \lambda_{12} Y^{(10)}_{1^{\prime}} \upsilon_{u}$,  $ \; \quad  \; $    $ a_{21} = \lambda_{21} Y^{(8)}_{1} \upsilon_{u}$, $ a_{32} = \lambda_{32} Y^{(8)}_{1} \upsilon_{u}$,  $ M_{N} = \Omega Y^{(4)}_{1} \phi$, $ \rho_{1}= \rho Y^{(8)}_{1^{\prime}} $ and $ \rho_{2} = \rho Y^{(10)}_{1}$.

\begin{center}
(d) ${\bf C_{3}}$ of Three-zero textures
\end{center}
\begin{eqnarray}
k_1=-3\;,\; k_2=-1\;, k_3=-1\;, \quad \\ \nonumber r_1=3 \;, r_2=1\;, r_3=1\;, \quad \\ \nonumber l_1 = -5\;,\; l_2 = -1\;,\; s = -5\;, k_{\phi}=-2.
\label{case d}
\end{eqnarray}
In this model of texture $ C_{3} $ of three-zeros, we utilize the multiplets of modular forms, namely  $Y^{(4)}_{1^{\prime}}$,  $Y^{(6)}_{1}$,$Y^{(8)}_{1}$,  $Y^{(8)}_{1^{\prime}}$, $Y^{(8)}_{1^{\prime \prime}}$, and $Y^{(10)}_{1^{\prime}}$ which were derived earlier. This results in a maximum value of $k_{Y{\rm max}} = 10$. The superpotential can be explicitly expressed as follows:
\begin{eqnarray}
W = &&(\alpha L^{}_1 H^{}_dE^c_1 +  \beta L^{}_2 H^{}_dE^c_2   + \gamma L^{}_3 H^{}_dE^c_3 ) 
\nonumber
\\
&& + (\lambda_{11} Y^{(8)}_{\bf 1^{\prime \prime}} L^{}_1 H^{}_u N^c_{1}  + \lambda_{12} Y^{(4)}_{\bf 1^{\prime}} L^{}_1 H^{}_u N^c_{2}  
\nonumber
\\
&& + \lambda_{21} Y^{(6)}_{\bf 1}  L^{}_2 H^{}_u N^c_{1}  + \lambda_{31} Y^{(8)}_{\bf 1^{\prime}}  L^{}_3H^{}_u N^c_{1})
\nonumber
\\
&&  + \Omega Y^{(8)}_{ \bf 1}\phi (N^c_{1}N^c_{2}+ N^c_{2}N^c_{1})
\\
&& + \rho( Y^{(10)}_{\textbf{1}^{\prime}} S N^c_{1} +  Y^{(6)}_{\textbf{1}}S N^c_{2}) \;\nonumber,
\label{casea:ed}
\end{eqnarray}
\begin{equation}
\resizebox{.46\textwidth}{!}
{$m_{\nu}=
\left(
\begin{matrix}
 \frac{2 a_{11} a_{12}}{M_N} & \frac{a_{12}^2}{M_N}+\frac{a_{11} a_{22}}{M_N} & \frac{a_{12} a_{31}}{M_N}+\frac{a_{11} a_{32}}{M_N} &\frac{a_{12} \rho_1}{M_N}+\frac{a_{11} \rho_2}{M_N}\\
 \frac{a_{12}^2}{M_N}+\frac{a_{11} a_{22}}{M_N} & 0 & 0 &  \frac{a_{22} \rho_1}{M_N}+\frac{a_{12} \rho_2}{M_N} \\
 \frac{a_{12} a_{31}}{M_N}+\frac{a_{11} a_{32}}{M_N} & 0 & 0 &\frac{a_{32} \rho_1}{M_N}+\frac{a_{31} \rho_2}{M_N} \\
  \frac{a_{12} \rho_1}{M_N}+\frac{a_{11} \rho_2}{M_N} & \frac{a_{22} \rho_1}{M_N}+\frac{a_{12} \rho_2}{M_N} & \frac{a_{32} \rho_1}{M_N}+\frac{a_{31} \rho_2}{M_N} &\frac{2 \rho_1 \rho_2}{M_N} 
\end{matrix}\right)$
}
\label{eqn:nu:d}
\end{equation}

where, $ a_{11} = \lambda_{11} Y^{(8)}_{1^{\prime \prime}} \upsilon_{u}$, $ a_{12} = \lambda_{12} Y^{(4)}_{1^{\prime}} \upsilon_{u}$,  $ \; \quad  \; $   $ a_{21} = \lambda_{21} Y^{(6)}_{1} \upsilon_{u}$, $ a_{31} = \lambda_{31} Y^{(8)}_{1^{\prime}} \upsilon_{u}$,  $ M_{N} = \Omega Y^{(8)}_{1}\phi$, $ \rho_{1}= \rho Y^{(10)}_{1^{\prime}} $ and $ \rho_{2} = \rho Y^{(6)}_{1}$.


\begin{center}
(e) ${\bf A_{3}}$ of Four-zero textures
\end{center}
\begin{eqnarray}
k_1=-1\; ,\; k_2=-3\;, k_3=1\;, \quad \\ \nonumber r_1=1 \; ,\; r_2=3\;, r_3=-1\;, \quad \\ \nonumber l_1 =-3\; ,\;  l_2=-5\;,\; s = -1, k_{\phi}=4\;,\;\;
\label{case e}
\end{eqnarray}
In this model of texture $ A_{3} $ of four-zeros, we utilize the multiplets of modular forms, namely $Y^{(4)}_{1}$, $Y^{(4)}_{1^{\prime}}$, $Y^{(6)}_{1}$, and $Y^{(8)}_{1^{\prime}}$, which were derived earlier. This results in a maximum value of $k_{Y{\rm max}} = 8$. The superpotential can be explicitly expressed as follows:
\begin{eqnarray}
W = &&(\alpha L^{}_1 H^{}_d E^c_1 +  \beta L^{}_2 H^{}_dE^c_2   + \gamma L^{}_3 H^{}_dE^c_3 ) 
\nonumber
\\
&& + (\lambda_{11} Y^{(4)}_{\bf 1} L^{}_1H^{}_u N^c_{1} + \lambda_{22} Y^{(8)}_{\bf 1^{\prime}}  L^{}_2H^{}_u N^c_{2}  
\nonumber
\\
&& + \lambda_{32} Y^{(4)}_{\bf 1}  L^{}_3H^{}_u N^c_{2} )
\nonumber
\\
&&  + \Omega Y^{(4)}_{ \bf 1}\phi (N^c_{1}N^c_{2}+ N^c_{2}N^c_{1})
\\
&& + \rho( Y^{(4)}_{\textbf{1}^{\prime}} S N^c_{1} +  Y^{(6)}_{\textbf{1}}S N^c_{2}) \;\nonumber,
\label{casea:ee}
\end{eqnarray}
\begin{equation}
\resizebox{.4\textwidth}{!}
{$m_{\nu}=
\left(
\begin{matrix}
0 & \frac{a_{12}^2}{M_N}+\frac{a_{11} a_{22}}{M_N} & \frac{a_{12} a_{31}}{M_N}+\frac{a_{11} a_{32}}{M_N} &\frac{a_{12} \rho_1}{M_N}+\frac{a_{11} \rho_2}{M_N}\\
\frac{a_{12}^2}{M_N}+\frac{a_{11} a_{22}}{M_N} & 0 & 0 &  \frac{a_{22} \rho_1}{M_N}+\frac{a_{12} \rho_2}{M_N} \\
 \frac{a_{12} a_{31}}{M_N}+\frac{a_{11} a_{32}}{M_N} & 0 & 0 &\frac{a_{32} \rho_1}{M_N}+\frac{a_{31} \rho_2}{M_N} \\
  \frac{a_{12} \rho_1}{M_N}+\frac{a_{11} \rho_2}{M_N} & \frac{a_{22} \rho_1}{M_N}+\frac{a_{12} \rho_2}{M_N} & \frac{a_{32} \rho_1}{M_N}+\frac{a_{31} \rho_2}{M_N} &\frac{2 \rho_1 \rho_2}{M_N} 
\end{matrix}\right)$
}
\label{eqn:nu:e}
\end{equation}
where, $ a_{11} = \lambda_{11} Y^{(4)}_{1} \upsilon_{u}$, $ a_{22} = \lambda_{22} Y^{(8)}_{1^{\prime}} \upsilon_{u}$,  $ \; \quad  \; $   $ a_{32} = \lambda_{32} Y^{(4)}_{1} \upsilon_{u}$,  $ M_{N} = \Omega Y^{(4)}_{1} \phi$, $ \rho_{1}= \rho Y^{(4)}_{1^{\prime}} $ $ \; \quad  \; $   and $ \rho_{2} = \rho Y^{(6)}_{1}$.

\begin{center}
(f) ${\bf A_{2}}$ of Four-zero textures
\end{center}
\begin{eqnarray}
k_1=1\;,\; k_2=-1\;, k_3=1\;, \quad \\ \nonumber r_1=-1 \;, r_2=1\;, r_3=-1\;, \quad \\ \nonumber l_1 = -5\;,\; l_2 = -3\;,\; s = -3\;, k_{\phi}=4. 
\label{case f}
\end{eqnarray}
In this model of texture $ A_{2} $ of four-zeros, we utilize the multiplets of modular forms, namely  $Y^{(4)}_{1}$,  $Y^{(4)}_{1^{\prime}}$,  $Y^{(6)}_{1}$, $Y^{(8)}_{1^{\prime}}$, and $Y^{(8)}_{1^{\prime \prime}}$, which were derived earlier. This results in a maximum value of $k_{Y{\rm max}} = 8$. The superpotential can be explicitly expressed as follows:
\begin{eqnarray}
W = &&(\alpha L^{}_1 H^{}_dE^c_1 +  \beta L^{}_2 H^{}_dE^c_2   + \gamma L^{}_3 H^{}_dE^c_3 ) 
\nonumber
\\
&& + (\lambda_{11} Y^{(4)}_{\bf 1} L^{}_1 H^{}_u N^c_{1}  + \lambda_{22} Y^{(4)}_{\bf 1^{\prime}} L^{}_2 H^{}_u N^c_{2}  
\nonumber
\\
&& + \lambda_{31} Y^{(4)}_{\bf 1^{\prime}}  L^{}_3 H^{}_u N^c_{1})
\nonumber
\\
&&  + \Omega Y^{(4)}_{ \bf 1} \phi (N^c_{1}N^c_{2}+ N^c_{2}N^c_{1})
\\
&& + \rho( Y^{(8)}_{\textbf{1}^{\prime}} S N^c_{1} +  Y^{(6)}_{\textbf{1}}S N^c_{2}) \;\nonumber,
\label{casea:ef}
\end{eqnarray}
\begin{equation}
\resizebox{.45\textwidth}{!}
{$m_{\nu}=
\left(
\begin{matrix}
 0 & \frac{a_{12}^2}{M_N}+\frac{a_{11} a_{22}}{M_N} & 0 &\frac{a_{12} \rho_1}{M_N}+\frac{a_{11} \rho_2}{M_N}\\
 \frac{a_{12}^2}{M_N}+\frac{a_{11} a_{22}}{M_N} & 0 & \frac{a_{22} a_{31}}{M_N}+\frac{a_{12} a_{32}}{M_N}&  \frac{a_{22} \rho_1}{M_N}+\frac{a_{12} \rho_2}{M_N} \\
 0 & \frac{a_{22} a_{31}}{M_N}+\frac{a_{12} a_{32}}{M_N} & 0 &\frac{a_{32} \rho_1}{M_N}+\frac{a_{31} \rho_2}{M_N} \\
  \frac{a_{12} \rho_1}{M_N}+\frac{a_{11} \rho_2}{M_N} & \frac{a_{22} \rho_1}{M_N}+\frac{a_{12} \rho_2}{M_N} & \frac{a_{32} \rho_1}{M_N}+\frac{a_{31} \rho_2}{M_N} &\frac{2 \rho_1 \rho_2}{M_N} 
\end{matrix}\right)$
}
\label{eqn:nu:f}
\end{equation}
where, $ a_{11} = \lambda_{11} Y^{(4)}_{1} \upsilon_{u}$, $ a_{22} = \lambda_{22} Y^{(4)}_{1^{\prime}} \upsilon_{u}$,  $ \; \quad  \; $   $ a_{31} = \lambda_{31} Y^{(4)}_{1^{\prime}} \upsilon_{u}$,  $ M_{N} = \Omega Y^{(4)}_{1}\phi$, $ \rho_{1}= \rho Y^{(8)}_{1^{\prime}} $ $ \; \quad  \; $   and $ \rho_{2} = \rho Y^{(6)}_{1}$.
\begin{figure}[h]  
     \centering  
   \includegraphics[width=0.45\textwidth]{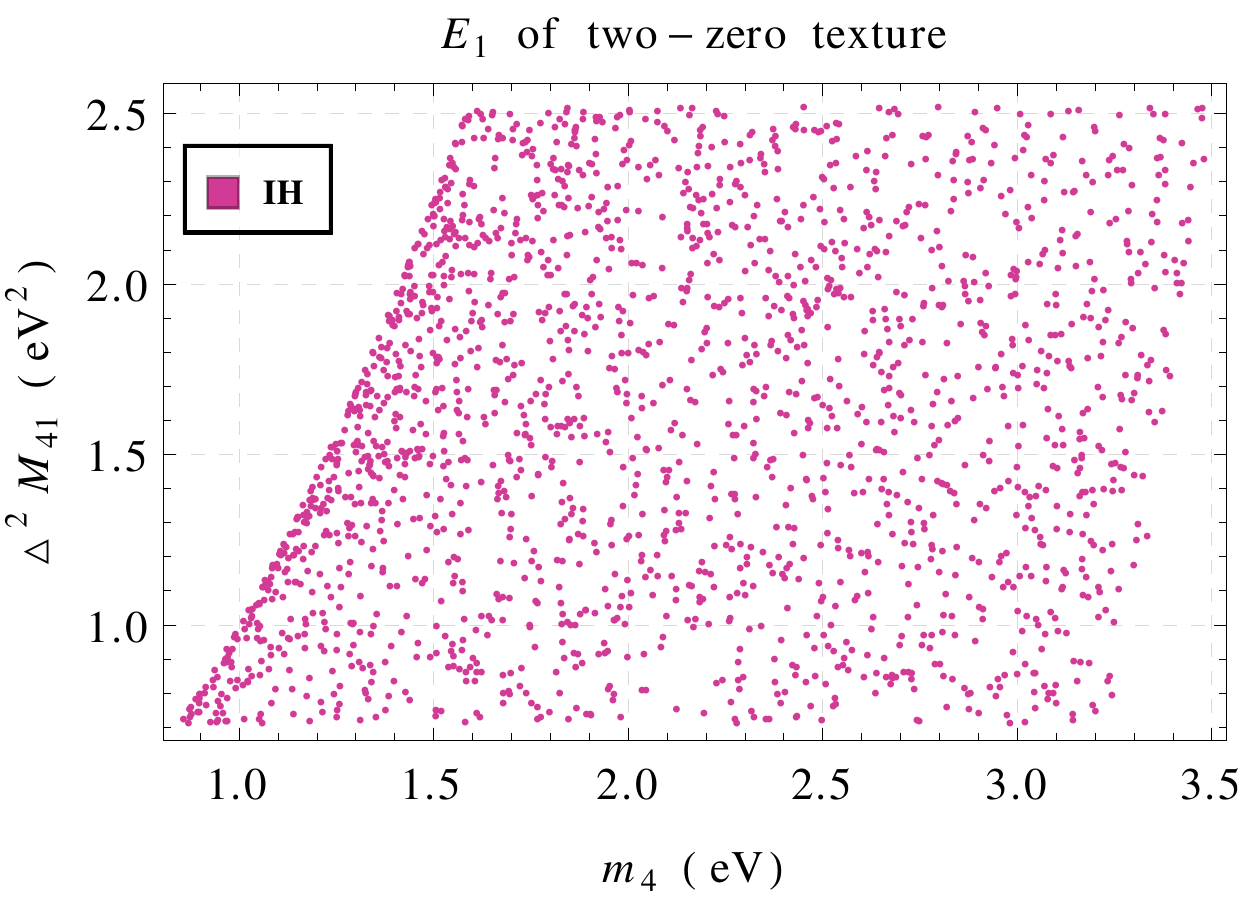} 
    \caption{}
         \label{Fig:2}
     \end{figure} 
     \begin{figure}[h]  
     \centering  
   \includegraphics[width=0.45\textwidth]{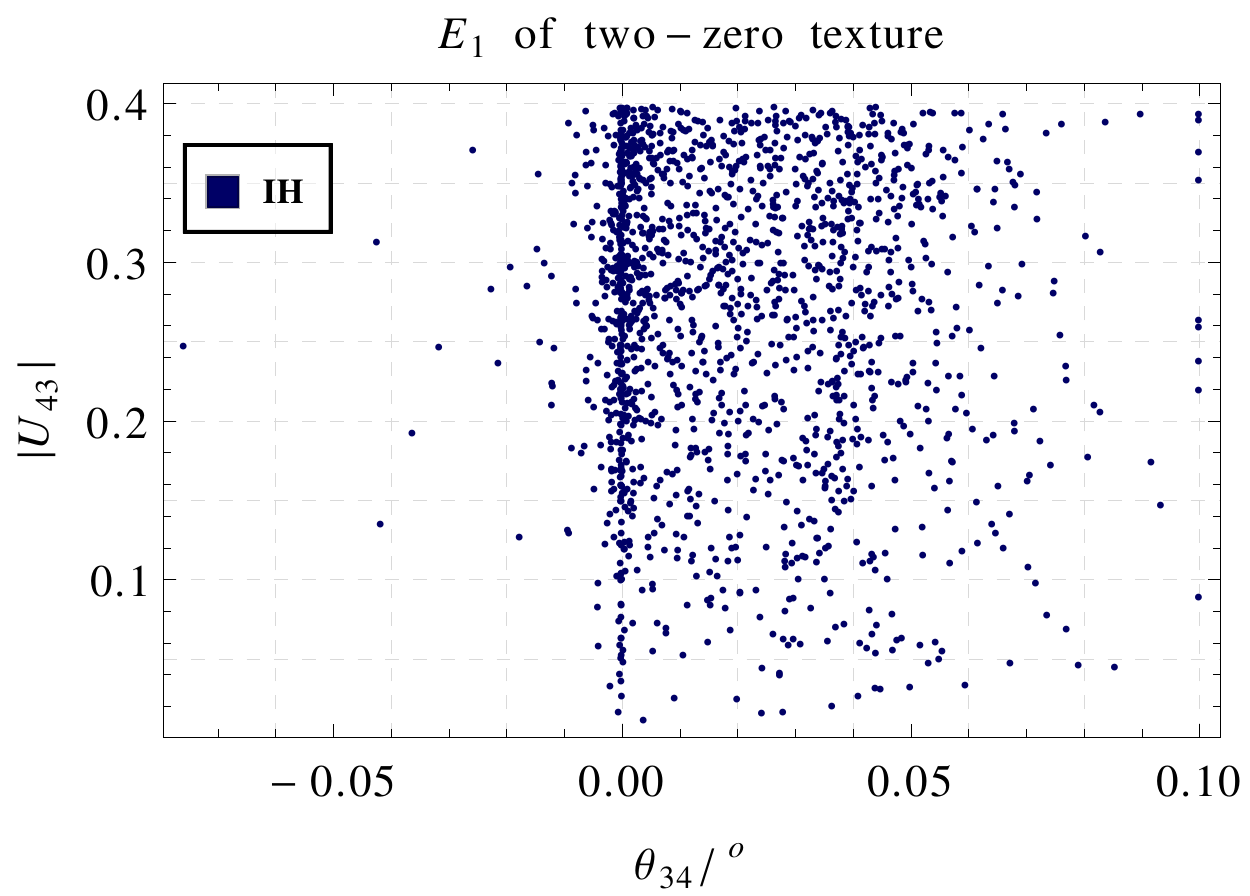} 
    \caption{}
         \label{Fig:3}
     \end{figure} 
     \begin{figure}[h]  
     \centering  
   \includegraphics[width=0.45\textwidth]{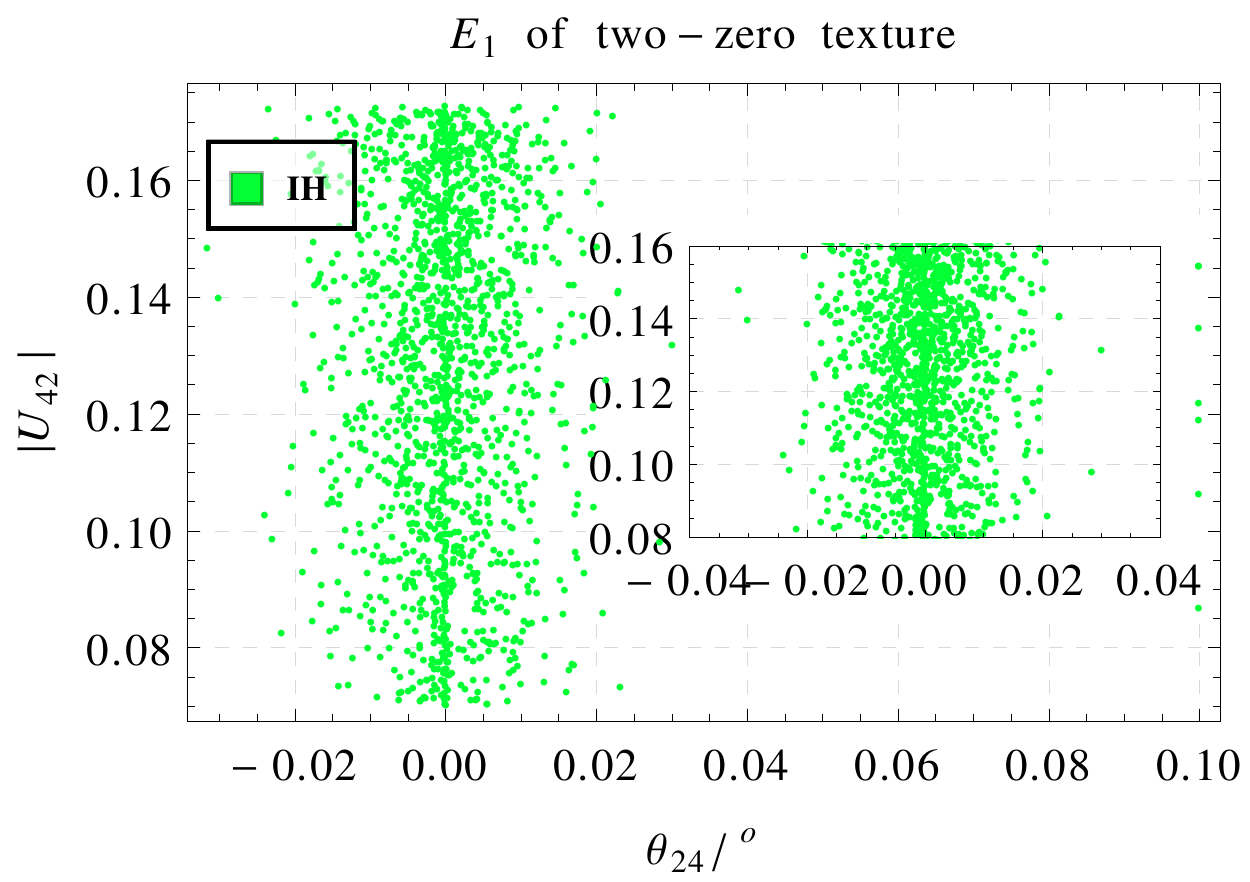} 
    \caption{}
         \label{Fig:4}
     \end{figure}
     \begin{figure}[h]  
     \centering  
   \includegraphics[width=0.45\textwidth]{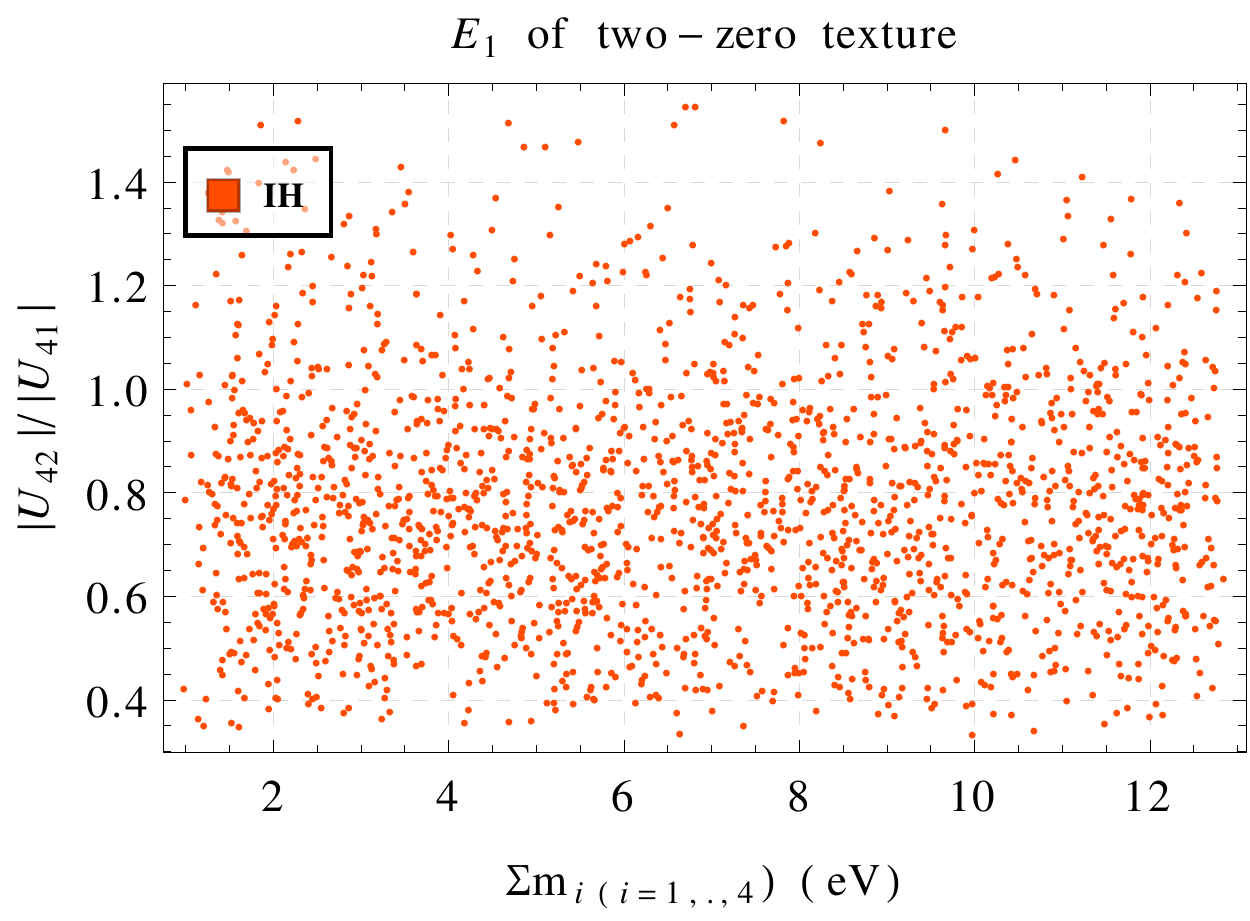} 
    \caption{}
         \label{Fig:5}
     \end{figure}
\section{Discussion and Summary}
Our research involves the implementation of the $A_{4}$ symmetry group and modular symmetry to identify the most promising textures from a range of possibilities, as outlined in \cite{Borah:2017azf}. To construct neutrino models, we assign $A_{4}$ representations to each chiral supermultiplet which are even weighted. This approach ensures that the Lagrangians remain invariant under modular transformation, satisfying the relation of $k_{Y}=k_{i}+k_{r}+...+ k_{l}$. The maximal weight $ k_{Y_{max}} $ for the neutrino models of the allowed cases varies from 4 to 10.
\\
\\
Our findings indicate that by fixing the texture of $M_{R}$ and $M_{S}$ matrices and varying only the texture of $M_{D}$, we can identify six allowed zero-texture of the active-sterile neutrino mixing matrix. These include three two-zero textures, one three-zero texture, and two four-zero textures. We have exclusively used here the minimal seesaw mechanism to derive the $4\times 4$ active-sterile neutrino mixing matrix.Our models have proven to be highly predictive and consistent with experimental values of active-sterile neutrino parameters within a 3$\sigma$ range. We have utilized the 3$\sigma$ values of both active and sterile neutrino oscillations, as shown in Table (\ref{tab:para}), using input variables such as $\Delta M^{2}_{41}$, $\theta_{14}$, $\theta_{24}$, and $\theta_{34}$ from recent publications \cite{Vien:2021hha}, \cite{Gariazzo:2017fdh}, \cite{Zhang:2013mb}, to determine unknown active and sterile neutrino parameters, such as $m_{lightest}$, $m_{4}$ (sterile neutrino mass), and Dirac CP violating phases $\delta_{14}$, $\delta_{24}$, $\delta_{13}$, as well as Majorana CPV phases $\alpha$ and $\beta$.
\\
\\
Our models have produced allowed textures that are highly consistent with experimental values, as mentioned in Table (\ref{tab:para}). For instance, if we consider the texture $ E_{1} $ texture of  the allowed two-zero textures of the active-sterile mixing matrix then it is found that the range of $ |U_{41} |^{2}$, $ |U_{24}|^{2} $, $ |U_{34}|^{2} $ are within the 3$ \sigma $ values of $ |U_{41} |^{2}$, $ |U_{24}|^{2} $, $ |U_{34}|^{2} $ as obtained in \cite{Vien:2021hha} for Inverted Hierarchy (IH). This is shown briefly in Table (\ref{tab:para1}). The correlation plots between  $m_{4}$ vs $ \Delta M^{2}_{41} $, $ \theta_{34}/^{o} $ vs $ |U_{43}| $, $ \theta_{24}/^{o} $ vs $ |U_{42}| $, and $ \Sigma m_{i (i=1,2,3,4)} $ vs $ |U_{42}|/|U_{41}| $,    for $ E_{1} $
two-zero texture is shown in Fig (\ref{Fig:2}), Fig(\ref{Fig:3}), Fig(\ref{Fig:4}) and Fig(\ref{Fig:5})respectively.
\begin{table}[h]
        \centering
         \scalebox{0.7}{
        \begin{tabular}{| c | c | c |}
        \hline 
\cline{1-3}
Parameters & Normal hierarchy & Inverted hierarchy \\ 
\hline
$ \Delta m^{2}_{21} $ $ (10^{-5} eV^{2}) $ & 6.95 $ \rightarrow $ 8.14 (7.50) & 6.94 $ \rightarrow $ 8.14 (7.50)  \\ 
\hline
$ |\Delta m^{2}_{31}| $ $ (10^{-3} eV^{2}) $ &  2.47 $ \rightarrow $ 2.63 (2.55) &  2.37 $ \rightarrow $ 2.53 (2.45)\\
 \hline
 $ Sin^{2} \theta^{2}_{12}/10^{-1}$& 2.71 $ \rightarrow $ 3.69 (3.18) & 2.71 $ \rightarrow $ 3.69 (3.18)\\
 \hline
 $ Sin^{2} \theta^{2}_{23}/10^{-1}$ & 4.34 $ \rightarrow $ 6.10 (5.74) & 4.33 $ \rightarrow $ 6.08 (5.78)\\
 \hline
$ Sin^{2} \theta^{2}_{13}/10^{-2}$ & 2.000 $ \rightarrow $ 2.405 (2.200) & 2.018 $ \rightarrow $ 2.424 (2.225)\\
 \hline
$ \delta/\pi $ & 0.71 $ \rightarrow $ 1.99 (1.08) & 1.11 $ \rightarrow $ 1.96 (1.58)\\
 \hline
 \hline
$ \Delta m^{2}_{41}/eV^{2} $  & 0.33 $ \rightarrow $ 148 & 0.33 $ \rightarrow $ 148\\
  \hline
 $ \theta_{14}/^{o} $  & 6.1 $ \rightarrow $ 15.3 & 6.1 $ \rightarrow $ 15.3\\
  \hline
  $ \theta_{24}/^{o} $  & 2.8 $ \rightarrow $ 12.5 & 2.8 $ \rightarrow $ 12.5\\
  \hline
  $ \theta_{34}/^{o} $  & 0 $ \rightarrow $ 30 & 0 $ \rightarrow $ 30\\
  \hline
$ |U_{14}|^{2} $ & 0.012 $ \rightarrow $ 0.047  & 0.012 $ \rightarrow $ 0.047\\
 \hline
$ |U_{24}|^{2} $ & 0.005 $ \rightarrow $ 0.03 & 0.005 $ \rightarrow $ 0.03\\
 \hline\
$ |U_{34}|^{2} $ & 0 $ \rightarrow $ 0.16 & 0 $ \rightarrow $ 0.16\\
 \hline
\end{tabular}}
      \caption{$ 3\sigma $ parameter values of active-sterile neutrinos \cite{Vien:2021hha}}
        \label{tab:para}
             \end{table}
             \begin{table}[h]
        \centering
         \scalebox{0.7}{
        \begin{tabular}{| c | c | c |}
        \hline 
        \multicolumn{3}{|c|}{Parameter values for $ E_{1} $ two zero texture in IH}\\
\cline{1-3}
Parameters  & Model predicted value & Experimental value ($ 3\sigma $)\\ 
\hline
$ |U_{14}|^{2} $ & 0.01202 $ \rightarrow $ 0.0469956  & 0.012 $ \rightarrow $ 0.047\\
 \hline
$ |U_{24}|^{2} $ & 0.00500654 $ \rightarrow $ 0.0299983 & 0.005 $ \rightarrow $ 0.03\\
 \hline\
$ |U_{34}|^{2} $ & 0.000185605 $ \rightarrow $ 0.159948 & 0 $ \rightarrow $ 0.16\\
 \hline
\end{tabular}}
      \caption{$ 3\sigma $ parameter values of $ |U_{14}|^{2} $, $ |U_{24}|^{2} $ and $ |U_{34}|^{2} $ as predicted for $ E_{1} $ two-zero texture model and latest $ 3\sigma $ value obtained from experiments.}
        \label{tab:para1}
             \end{table}
\\
\\
Overall, our research provides valuable insights into the use of symmetry groups and modular symmetry to construct neutrino models. By identifying the most viable textures of active-sterile neutrino, we can better understand the behavior of active-sterile neutrino mixing matrices and their implication in Dark Matter, Baryon Asymmetry of the Universe etc.\\
\\
\textbf{Acknowledgements}\\
\\
The author expresses sincere gratitude to Xiang-Gan Liu from UC Irvine in USA, Prof. Ferruccio Feruglio from INFN in Italy, Shivajee Gupta from IITGN, Monal Kashav from Central University of Himachal Pradesh and Prof. Ramakrishnan Balakrishnan from the Indian Statistical Institute (TASU-NE) for their invaluable guidance, important inputs and insightful discussions on modular forms. Their expertise has been instrumental in shaping the author's understanding of this complex topic. Additionally, the author would like to extend thanks to Prof. Rupam Barman from the Department of Mathematics at IIT Guwahati for granting permission to attend the NCWM-ATM Schools on Jacobi Forms. The lectures delivered during the workshop have been an immense help in comprehending the intricate concepts of Modular and Jacobi Forms. Overall, the author is deeply appreciative of the support and knowledge provided by these esteemed individuals, which has undoubtedly contributed to the success of this research work.
\vspace{-10px}
\appendix
\section{The Tensor Product and Clebsch Gordon coefficient of $ A_{4} $ symmetry}
\label{appen2}
$ A_{4} $ group \cite{Ma:2001dn}, \cite{Babu:2002dz}, \cite{Altarelli:2005yp}, \cite{Altarelli:2005yp}, \cite{Ishimori:2010au} is an alternating group (of even permutation) of object 4. It has two generators S and T which represents rotation and translation. The tensor product and Clebsch Gordon coefficient for $ A_{4} $ symmmetry is shown below:

The $ A_{4} $ group has a total of 12 elements that originates from 4 conjugacy classes. It has  three singlets and one triplet irreducible representation
\\
\\
The product rules of $ A_{4} $ symmetry is given by

\begin{eqnarray}
{\bf 1^\prime} \otimes  {\bf 1^\prime} =  {\bf 1^{\prime\prime}} \;,\quad 
{\bf 1^{\prime\prime}} \otimes   {\bf 1^{\prime\prime}} = {\bf 1^\prime} \;,\quad \nonumber\\
{\bf 1^\prime} \otimes  {\bf 1^{\prime\prime}} = {\bf 1^{\prime\prime}} \otimes  {\bf 1^\prime} =  {\bf 1},  \;\;\quad \;\;\\ {\bf 3} \otimes  {\bf 3} = {\bf 1} \oplus {\bf 1^\prime} \oplus {\bf 1^{\prime\prime}} \oplus {\bf 3_{s}} \oplus {\bf 3_{a}} \;\quad. \nonumber
\end{eqnarray}

\section{The allowed textures of 3+1 Majorana mass matrix under $ A_{4} $ modular group}
\label{appen1}
\begin{table}[h]
        \centering
         \scalebox{0.5}{
        \begin{tabular}{| c | c | c | c |}
        \hline 
\cline{1-4}
\cellcolor{gray!20}$ A_{1} $ & \cellcolor{gray!20} $ A_{2} $ &   &  \\ 
\hline
\cellcolor{gray!8}$ \left(\begin{matrix}
 0 & 0 & \times & \times\\
0 & \times & \times & \times\\
\times & \times & \times &\times\\
\times & \times & \times &\times\\
\end{matrix}\right) $ & \cellcolor{gray!8} $ \left(\begin{matrix}
0 & \times & 0 & \times\\
\times & \times & \times & \times\\
0 & \times & \times &\times\\
\times & \times & \times &\times\\
\end{matrix}\right) $ &  & \\
\hline
\cellcolor{gray!20}$ B_{1} $ & \cellcolor{gray!20} $ B_{2} $ &\cellcolor{gray!20} $ B_{3} $ &\cellcolor{gray!20} $ B_{4} $ \\
 \hline
\cellcolor{gray!8} $ \left(\begin{matrix}
\times & \times & 0 & \times\\
\times & 0 & \times & \times\\
0 & \times & \times &\times\\
\times & \times & \times &\times\\
\end{matrix}\right) $ & \cellcolor{gray!8} $ \left(\begin{matrix}
\times & 0 & \times & \times\\
0 & \times & \times & \times\\
\times & \times & 0 &\times\\
\times & \times & \times &\times\\
\end{matrix}\right) $  &  \cellcolor{gray!8}$ \left(\begin{matrix}
\times & 0 & \times & \times\\
0 & 0 & \times & \times\\
\times & \times & \times &\times\\
\times & \times & \times &\times\\
\end{matrix}\right) $ &  \cellcolor{gray!8} $ \left(\begin{matrix}
\times & \times & 0 & \times\\
\times & \times & \times & \times\\
0 & \times & 0 &\times\\
\times & \times & \times &\times\\
\end{matrix}\right) $ \\
 \hline
\cellcolor{gray!20} C &  &  &  \\
 \hline
\cellcolor{gray!8}$ \left(\begin{matrix}
\times & \times & \times & \times\\
\times & 0 & \times & \times\\
\times & \times & 0 &\times\\
\times & \times & \times &\times\\
\end{matrix}\right) $  &  &  &  \\
\hline 
\cellcolor{gray!20} $ D_{1} $ &\cellcolor{gray!20} $ D_{2} $ &  &  \\
 \hline
 \cellcolor{gray!8} $ \left(\begin{matrix}
\times & \times & \times & \times\\
\times & 0 & 0 & \times\\
\times & 0 & \times &\times\\
\times & \times & \times &\times\\
\end{matrix}\right) $ & \cellcolor{gray!8} $ \left(\begin{matrix}
\times & \times & \times & \times\\
\times & \times & 0 & \times\\
\times & 0 & 0 &\times\\
\times & \times & \times &\times\\
\end{matrix}\right) $  &  &  \\
\hline 
\cellcolor{gray!20} $  E_{1}$ &\cellcolor{gray!20} $ E_{2} $  &\cellcolor{gray!20} $ E_{3} $  &  \\
 \hline
\cellcolor{gray!8} $ \left(\begin{matrix}
0 & \times & \times & \times\\
\times & 0 & \times & \times\\
\times & \times & \times &\times\\
\times & \times & \times &\times\\
\end{matrix}\right) $  & \cellcolor{gray!8} $ \left(\begin{matrix}
0 & \times & \times & \times\\
\times & \times & \times & \times\\
\times & \times & 0 &\times\\
\times & \times & \times &\times\\
\end{matrix}\right) $   &\cellcolor{gray!8} $ \left(\begin{matrix}
0 & \times & \times & \times\\
\times & \times & 0 & \times\\
\times & 0 & \times &\times\\
\times & \times & \times &\times\\
\end{matrix}\right) $  &  \\
\hline 
\cellcolor{gray!20} $  F_{1}$ &\cellcolor{gray!20} $ F_{2} $  &\cellcolor{gray!20} $ F_{3} $ &  \\
 \hline
\cellcolor{gray!8} $ \left(\begin{matrix}
\times & 0 & 0 & \times\\
0 & \times & \times & \times\\
0 & \times & \times &\times\\
\times & \times & \times &\times\\
\end{matrix}\right) $  & \cellcolor{gray!8} $ \left(\begin{matrix}
\times & 0 & \times & \times\\
0 & \times & 0 & \times\\
\times & 0 & \times &\times\\
\times & \times & \times &\times\\
\end{matrix}\right) $  & \cellcolor{gray!8} $ \left(\begin{matrix}
\times & \times & 0 & \times\\
\times & \times & 0 & \times\\
0 & 0 & \times &\times\\
\times & \times & \times &\times\\
\end{matrix}\right) $   &  \\
\hline 
\end{tabular}}
      \caption{Possible two-zero textures in $ m_{\nu} $ in the 3+1 scenario}
        \label{tab:2-0}
             \end{table}

\begin{table}[h]
        \centering
         \scalebox{0.55}{
        \begin{tabular}{| c | c | c |}
        \hline 
\cline{1-3}
\cellcolor{gray!20}$ A $ &  &   \\ 
\hline
\cellcolor{gray!8}$ \left(\begin{matrix}
 0 & \times & \times & \times\\
\times & 0 & \times & \times\\
\times & \times & 0 &\times\\
\times & \times & \times &\times\\
\end{matrix}\right) $ &  &  \\
\hline
\cellcolor{gray!20}$ B_{1} $ & \cellcolor{gray!20} $ B_{2} $ &\cellcolor{gray!20} $ B_{3} $  \\
 \hline
\cellcolor{gray!8} $ \left(\begin{matrix}
0 & \times & 0 & \times\\
\times & 0 & \times & \times\\
0 & \times & \times &\times\\
\times & \times & \times &\times\\
\end{matrix}\right) $ & \cellcolor{gray!8} $ \left(\begin{matrix}
0 & \times & \times & \times\\
\times & 0 & 0 & \times\\
\times & 0 & \times &\times\\
\times & \times & \times &\times\\
\end{matrix}\right) $  &  \cellcolor{gray!8}$ \left(\begin{matrix}
0 & 0 & \times & \times\\
0 & \times & \times & \times\\
\times & \times & 0 &\times\\
\times & \times & \times &\times\\
\end{matrix}\right) $ \\
 \hline
\cellcolor{gray!20} $B_{4}$ & \cellcolor{gray!20}$B_{5}$ &\cellcolor{gray!20} $B_{6}$  \\
 \hline
\cellcolor{gray!8}$ \left(\begin{matrix}
0 & \times & \times & \times\\
\times & \times & 0 & \times\\
\times & 0 & 0 &\times\\
\times & \times & \times &\times\\
\end{matrix}\right) $  & \cellcolor{gray!8}$ \left(\begin{matrix}
\times & 0 & \times & \times\\
0 & 0 & \times & \times\\
\times & \times & 0 &\times\\
\times & \times & \times &\times\\
\end{matrix}\right) $  & \cellcolor{gray!8}$ \left(\begin{matrix}
\times & \times & 0 & \times\\
\times & 0 & \times & \times\\
0 & \times & 0 &\times\\
\times & \times & \times &\times\\
\end{matrix}\right) $    \\
\hline 
\cellcolor{gray!20} $ C_{1} $ &\cellcolor{gray!20} $ C_{2} $ &   \cellcolor{gray!20} $ C_{3} $  \\
 \hline
 \cellcolor{gray!8} $ \left(\begin{matrix}
0 & 0 & \times & \times\\
0 & 0 & \times & \times\\
\times & \times & \times & \times\\
\times & \times & \times & \times\\
\end{matrix}\right) $ & \cellcolor{gray!8} $ \left(\begin{matrix}
0 & \times & 0 & \times\\
\times & \times & \times & \times\\
0 & \times & 0 &\times\\
\times & \times & \times &\times\\
\end{matrix}\right) $  &  \cellcolor{gray!8} $ \left(\begin{matrix}
\times & \times & \times & \times\\
\times & 0 & 0 & \times\\
\times & 0 & 0 &\times\\
\times & \times & \times &\times\\
\end{matrix}\right) $  \\
\hline 
\cellcolor{gray!20} $  D_{1}$ &\cellcolor{gray!20} $ D_{2} $  &\cellcolor{gray!20} $ D_{3} $    \\
 \hline
\cellcolor{gray!8} $ \left(\begin{matrix}
0 & 0 & \times & \times\\
0 & \times & 0 & \times\\
\times & 0 & \times &\times\\
\times & \times & \times &\times\\
\end{matrix}\right) $  & \cellcolor{gray!8} $ \left(\begin{matrix}
0 & \times & 0 & \times\\
\times & \times & 0 & \times\\
0 & 0 & \times &\times\\
\times & \times & \times &\times\\
\end{matrix}\right) $   & \cellcolor{gray!8} $ \left(\begin{matrix}
\times & 0 & 0 & \times\\
0 & 0 & \times & \times\\
0 & \times & \times &\times\\
\times & \times & \times &\times\\
\end{matrix}\right) $ \\
\hline 
\cellcolor{gray!20} $  D_{4}$ &\cellcolor{gray!20} $ D_{5} $  &\cellcolor{gray!20} $ D_{6} $  \\
 \hline
\cellcolor{gray!8} $ \left(\begin{matrix}
\times & \times & 0 & \times\\
\times & 0 & 0 & \times\\
0 & 0 & \times &\times\\
\times & \times & \times &\times\\
\end{matrix}\right) $  & \cellcolor{gray!8} $ \left(\begin{matrix}
\times & 0 & 0 & \times\\
0 & \times & \times & \times\\
0 & \times & 0 &\times\\
\times & \times & \times &\times\\
\end{matrix}\right) $  & \cellcolor{gray!8} $ \left(\begin{matrix}
\times & 0 & \times & \times\\
0 & \times & 0 & \times\\
\times & 0 & 0 &\times\\
\times & \times & \times &\times\\
\end{matrix}\right) $    \\
\hline 
\cellcolor{gray!20} $  E_{1}$ &\cellcolor{gray!20} $ E_{2} $  &\cellcolor{gray!20} $ E_{3} $  \\
 \hline
\cellcolor{gray!8} $ \left(\begin{matrix}
0 & 0 & 0 & \times\\
0 & \times & \times & \times\\
0 & \times & \times &\times\\
\times & \times & \times &\times\\
\end{matrix}\right) $  & \cellcolor{gray!8} $ \left(\begin{matrix}
\times & 0 & \times & \times\\
0 & 0 & 0 & \times\\
\times & 0 & \times &\times\\
\times & \times & \times &\times\\
\end{matrix}\right) $  & \cellcolor{gray!8} $ \left(\begin{matrix}
\times & \times & 0 & \times\\
\times & \times & 0 & \times\\
0 & 0 & 0 &\times\\
\times & \times & \times &\times\\
\end{matrix}\right) $    \\
\hline 
\end{tabular}}
      \caption{Possible three-zero textures in $ m_{\nu} $ in the 3+1 scenario}
        \label{tab:3-0}
             \end{table}
             
             \begin{table}[h]
        \centering
         \scalebox{0.55}{
        \begin{tabular}{| c | c | c |}
        \hline 
\cline{1-3}
\cellcolor{gray!20}$ A_{1} $ &\cellcolor{gray!20} $ A_{2} $ & \cellcolor{gray!20}$ A_{3} $   \\ 
\hline
\cellcolor{gray!8}$ \left(\begin{matrix}
 0 & 0 & \times & \times\\
0 & 0 & \times & \times\\
\times & \times & 0 &\times\\
\times & \times & \times &\times\\
\end{matrix}\right) $ & \cellcolor{gray!8}$ \left(\begin{matrix}
 0 & \times & 0 & \times\\
\times & 0 & \times & \times\\
0 & \times & 0 &\times\\
\times & \times & \times &\times\\
\end{matrix}\right) $ & \cellcolor{gray!8}$ \left(\begin{matrix}
 0 & \times & \times & \times\\
\times & 0 & 0 & \times\\
\times & 0 & 0 &\times\\
\times & \times & \times &\times\\
\end{matrix}\right) $ \\
\hline
\cellcolor{gray!20}$ A_{4} $ & \cellcolor{gray!20} $ A_{5} $ &\cellcolor{gray!20} $ A_{6} $  \\
 \hline
\cellcolor{gray!8} $ \left(\begin{matrix}
0 & 0 & 0 & \times\\
0 & 0 & \times & \times\\
0 & \times & \times &\times\\
\times & \times & \times &\times\\
\end{matrix}\right) $ & \cellcolor{gray!8} $ \left(\begin{matrix}
0 & 0 & \times & \times\\
0 & 0 & 0 & \times\\
\times & 0 & \times &\times\\
\times & \times & \times &\times\\
\end{matrix}\right) $  &  \cellcolor{gray!8}$ \left(\begin{matrix}
0 & \times & 0 & \times\\
\times & 0 & 0 & \times\\
0 & 0 & \times &\times\\
\times & \times & \times &\times\\
\end{matrix}\right) $ \\
 \hline
\cellcolor{gray!20} $A_{7}$ & \cellcolor{gray!20}$A_{8}$ & \cellcolor{gray!20}$A_{9}$  \\
 \hline
\cellcolor{gray!8}$ \left(\begin{matrix}
0 & 0 & 0 & \times\\
0 & \times & \times & \times\\
0 & \times & 0 &\times\\
\times & \times & \times &\times\\
\end{matrix}\right) $  & \cellcolor{gray!8}$ \left(\begin{matrix}
0 & 0 & \times & \times\\
0 & \times & 0 & \times\\
\times & 0 & 0 &\times\\
\times & \times & \times &\times\\
\end{matrix}\right) $  & \cellcolor{gray!8}$ \left(\begin{matrix}
0 & \times & 0 & \times\\
\times & \times & 0 & \times\\
0 & 0 & 0 &\times\\
\times & \times & \times &\times\\
\end{matrix}\right) $    \\
\hline 
\cellcolor{gray!20} $ A_{10} $ &\cellcolor{gray!20} $ B_{1} $ &   \cellcolor{gray!20} $ B_{2} $  \\
 \hline
 \cellcolor{gray!8} $ \left(\begin{matrix}
0 & 0 & 0 & \times\\
0 & \times & 0 & \times\\
0 & 0 & \times & \times\\
\times & \times & \times & \times\\
\end{matrix}\right) $ & \cellcolor{gray!8} $ \left(\begin{matrix}
\times & 0 & \times & \times\\
0 & 0 & 0 & \times\\
\times & 0 & 0 &\times\\
\times & \times & \times &\times\\
\end{matrix}\right) $  &  \cellcolor{gray!8} $ \left(\begin{matrix}
\times & \times & 0 & \times\\
\times & 0 & 0 & \times\\
0 & 0 & 0 &\times\\
\times & \times & \times &\times\\
\end{matrix}\right) $  \\
\hline 
\cellcolor{gray!20} $  B_{3}$ & \cellcolor{gray!20} $ B_{4} $  & \cellcolor{gray!20} $ B_{5} $    \\
 \hline
\cellcolor{gray!8} $ \left(\begin{matrix}
\times & 0 & 0 & \times\\
0 & 0 & \times & \times\\
0 & 0 & 0 &\times\\
\times & \times & \times &\times\\
\end{matrix}\right) $  & \cellcolor{gray!8} $ \left(\begin{matrix}
\times & 0 & 0 & \times\\
0 & 0 & 0 & \times\\
0 & 0 & \times &\times\\
\times & \times & \times &\times\\
\end{matrix}\right) $   & \cellcolor{gray!8} $ \left(\begin{matrix}
\times & 0 & 0 & \times\\
0 & \times & 0 & \times\\
0 & 0 & 0 &\times\\
\times & \times & \times &\times\\
\end{matrix}\right) $ \\
\hline 
\end{tabular}}
      \caption{Possible four-zero textures in $ m_{\nu} $ in the 3+1 scenario}
        \label{tab:4-0}
             \end{table}
             
                       \begin{table}[h]
        \centering
         \scalebox{0.55}{
        \begin{tabular}{| c | c | c |}
        \hline  
\cline{1-3}
\cellcolor{gray!20}$ A $ &\cellcolor{gray!20} $ B $ & \cellcolor{gray!20} $ C $   \\ 
\hline
\cellcolor{gray!8}$ \left(\begin{matrix}
 0 & 0 & 0 & \times\\
0 & 0 & 0 & \times\\
0 & 0 & \times &\times\\
\times & \times & \times &\times\\
\end{matrix}\right) $ & \cellcolor{gray!8}$ \left(\begin{matrix}
 0 & 0 & 0 & \times\\
0 & 0 & \times & \times\\
0 & \times & 0 &\times\\
\times & \times & \times &\times\\
\end{matrix}\right) $ & \cellcolor{gray!8}$ \left(\begin{matrix}
 0 & 0 & 0 & \times\\
0 & \times & 0 & \times\\
0 & 0 & 0 &\times\\
\times & \times & \times &\times\\
\end{matrix}\right) $ \\
\hline
\cellcolor{gray!20}$ D $ & \cellcolor{gray!20} $ E $ &\cellcolor{gray!20} $ F $  \\
 \hline
\cellcolor{gray!8} $ \left(\begin{matrix}
0 & 0 & \times & \times\\
0 & 0 & 0 & \times\\
\times & 0 & 0 &\times\\
\times & \times & \times &\times\\
\end{matrix}\right) $ & \cellcolor{gray!8} $ \left(\begin{matrix}
0 & \times & 0 & \times\\
\times & 0 & 0 & \times\\
0 & 0 & 0 &\times\\
\times & \times & \times &\times\\
\end{matrix}\right) $  &  \cellcolor{gray!8}$ \left(\begin{matrix}
\times & 0 & 0 & \times\\
0 & 0 & 0 & \times\\
0 & 0 & 0 &\times\\
\times & \times & \times &\times\\
\end{matrix}\right) $ \\
\hline 
\end{tabular}}
      \caption{Possible five-zero textures in $ m_{\nu} $ in the 3+1 scenario}
        \label{tab:5-0}
             \end{table}
This section provides a concise summary of the permitted zero-texture 3+1 neutrino mass matrices, which have been extensively reviewed and elaborately explained in the reference \cite{Borah:2017azf}. When considering the matrix of active-sterile neutrino mixing with the sterile neutrino as the fourth generation, there are a total of 45 allowed two-zero textures in $ m_{\nu} $. However, only 15 of these two-zero textures are actually possible. Interestingly, these 15 allowed textures for active-sterile neutrino mixing are the same as the 15 allowed textures of the neutrino mass matrix with only three generations. To provide further clarity, we have included a Table (\ref{tab:2-0}) displaying the 15 permitted two-zero textures.\\ 
\\
When dealing with three-zero textures, there are 120 possible textures to consider. However, only 20 of these textures are allowed, as the other 100 have been discarded. These 20 textures can be found in Table (\ref{tab:3-0}). It's worth noting that the $ C_{3} $ texture is only allowed in a very small region of the IH parameter space as mentioned in the reference \cite{Borah:2017azf}.
\\
\\
Moving on to four-zero textures, there are a total of 210 textures to analyze. Unfortunately, 195 of these textures have been discarded, leaving only 15 allowed textures. These 15 textures can be found in Table (\ref{tab:4-0}).
\\
\\
When it comes to five-zero textures, there are 252 possible options. However, only six of these textures are allowed. It's important to note that none of our allowed $ m_{\nu} $ textures fall under the category of five-zero textures. The possible textures for five-zero can be found in Table (\ref{tab:5-0}).
\\
\\
In summary, while there are many possible textures to consider, only few of them are allowed. By analyzing these allowed textures with mathematical tools such as modular and flavour symmetries, we can gain a better understanding of 3+1 neutrino mixing matrix.

\end{document}